\documentclass[a4paper,11pt]{article}
\pdfoutput=1

\usepackage{amssymb,amsmath,amsfonts,lmodern}
\usepackage{amsthm}
\usepackage{float}
\usepackage{graphicx}
\usepackage{hyperref}
\usepackage{mathtools}
\usepackage{subcaption}

\usepackage{tikz}
\usetikzlibrary{matrix,shapes, arrows, shadows}

\tikzstyle{vertex} = [circle, draw, fill=blue!20, scale=1,auto=left]
\tikzstyle{vert} = [circle, draw, fill=blue!20, scale=.8,auto=left]
\tikzstyle{line} = [draw]

\newcommand{\midarrow}{\tikz \draw[-triangle 90] (0,0) -- +(.1,0);}

\usepackage{cite}

\usepackage{upgreek}
\usepackage{braket}

\def\be{\begin{equation}}
\def\ee{\end{equation}}
\def\bea{\begin{eqnarray}}
\def\eea{\end{eqnarray}}

\begin{document}

\begin{titlepage}
\date{\today}       \hfill

\begin{center}

\vskip .5in

{\LARGE \bf   On asymptotic behaviour in truncated conformal space approach }\\
\vspace{5mm}

\today
 
\vskip .250in

\vskip .5in
{\large Anatoly Konechny  and Dermot McAteer}

\vskip 0.5cm
{\it Department of Mathematics,  Heriot-Watt University\\
Edinburgh EH14 4AS, United Kingdom\\[10pt]
and \\[10pt]
Maxwell Institute for Mathematical Sciences\\
Edinburgh, United Kingdom\\[10pt]
}
E-mail: A.Konechny@hw.ac.uk, dm77@hw.ac.uk
\end{center}

\vskip .5in
\begin{abstract} \large
The Truncated conformal space approach (TCSA) is a numerical technique for finding finite size spectrum of 
Hamiltonians in quantum field theory described as perturbations of conformal field theories. 
The truncation errors of the method have been systematically studied near the UV fixed point 
(when  the characteristic energy related to the coupling is less than the truncation cutoff) where a good theoretical understanding has been achieved. 
However numerically the method demonstrated a good agreement with other methods 
for much larger values of the coupling when the RG flow approaches a new fixed point in the infrared. 
In the present paper we investigate this regime for a number of boundary RG flows testing the leading exponent and 
truncation errors. We also study the flows beyond the 
first fixed point which have been  observed numerically but yet lack a theoretical understanding. We show that while in some 
models such flows approximate reversed physical RG flows, in other models the spectrum approaches a stable 
regime that does not correspond to any local boundary condition. Furthermore we find that in general the flows beyond the first fixed point are 
very sensitive to modifications of the truncation scheme. 
\end{abstract}

\end{titlepage}

\renewcommand{\thepage}{\arabic{page}}
\setcounter{page}{1}
\large 

\section{Introduction }
\renewcommand{\theequation}{\arabic{section}.\arabic{equation}}
The Truncated conformal space approach (TCSA) was invented in \cite{YZ1}, \cite{YZ2} to study two dimensional conformal 
field theories (2D CFTs) perturbed by a relevant operator. TCSA proved to be a reliable numerical method for 
studying the strong-coupling physics of such quantum field theories.   
For the majority of these, the large coupling limit is described by  a trivial theory in which only a vacuum 
(possibly degenerate) state survives in the low energy sector.  Examples of RG flows 
ending in a non-trivial infrared fixed point studied by TCSA are relatively scarce. The situation is different for boundary RG flows 
which always end up in a non-trivial fixed point. In the present paper we limit ourselves to studying large coupling behaviour 
in TCSA numerics for perturbed boundary conformal field theories (BCFT).

TCSA was adopted to boundary perturbations  in \cite{DPTW}. In this case we consider a 2D CFT on a strip of width $L$ with a choice of 
conformal boundary condition at each end.  Let $0\le \sigma \le L$ be the coordinate across the strip and $-\infty <\tau<\infty$ be the coordinate 
along the strip and let $s$ and $s'$ be the boundary condition labels for $\sigma=0$ and $\sigma=L$ respectively. 
Choosing $\tau$ to be Euclidean time gives a quantisation scheme in which the Hilbert space ${\cal H}^{(s,s')}$ for a theory on an interval 
of length $L$ splits into Virasoro irreducible representations ${\cal H}_{i}$. For a diagonal Virasoro minimal model 
$s$ and $s'$ are just primary state labels and   
the state space decomposition has the form \cite{Cardy}
\be \label{Hss}
{\cal H}^{(s,s')} = \bigoplus_{i} N_{s, s'}^{i}{\cal H}_{i}
\ee
where $N_{s, s'}^{i}$ are the fusion coefficients. Perturbing the boundary condition on the $\sigma=0$ end with a boundary 
operator $\psi$ is described by a Hamiltonian 
\be \label{H_pert_gen}
H=\frac{\pi}{L}[ L_{0} - \frac{c}{24} + L\mu \psi(0,0) ]
\ee
where $\mu$ is a dimensionful coupling. 

 States in ${\cal H}_{i}$ are obtained as Virasoro descendants of the primary field $|\psi_{i}\rangle$. They are 
 spanned by states of the form 
 \be \label{desc_states}
 L_{-k_1}\dots L_{-k_{n}}|\psi_{i}\rangle \qquad \enspace k_{j} > 0
 \ee
 where $L_{-k_j}$ are Virasoro modes. In practice it is convenient to work with a linear basis, not necessarily orthonormal, 
  of states of the form (\ref{desc_states}). For minimal models, where we have null states, 
   we choose such a basis for a subspace on which the Gramm's matrix is non-degenerate. 
    
 In TCSA we truncate the state space  (\ref{Hss}) keeping 
only linear combinations of states whose   conformal weight is smaller than some maximal weight set by a truncation parameter $\Delta_{\rm max}$. 
In the simplest form we can take all weights that satisfy
\be \label{simple_prescription}
\Delta \le \Delta_{\rm max}\, . 
\ee
 Here the Virasoro weights $\Delta$ are the eigenvalues of $L_{0}$ and are of the form $\Delta = h_{i} + N$ where $h_{i}$ is the value of the primary weight in the given Virasoro irreducible 
representation and $N$ is a non-negative  integer that gives the weight of the Virasoro descendants.  
For a state of the form (\ref{desc_states}) $N=\sum_{j=1}^{n} k_{j}$. 
 Assume for simplicity that we have 
finitely many primaries labelled by $i=1,\dots, p$.  Depending on the value of $\Delta_{\rm max}$ 
the prescription  (\ref{simple_prescription}) sets the maximum descendant level $N$ for each conformal tower labelled by the primary weight $h_i$. Thus, 
given $\Delta_{\rm max}$ we get a number of bounds $N_{\rm max}(h_i, \Delta_{\rm max})$. A more general truncation scheme, still based only on conformal weights, is specified by setting  independent maximal descendant levels $(N_{\rm max}^{(1)}, \dots, N_{\rm max}^{(p)} )$ so that we keep states 
with weights $\Delta = h_{i} + N_{i}$ in the corresponding conformal towers satisfying 
\be \label{modif_scheme}
N_{1} \le N_{\rm max}^{(1)} \, , \enspace N_{2} \le N_{\rm max}^{(2)}\, , \dots \, ,  N_{p} \le N_{\rm max}^{(p)} \, . 
\ee

The perturbed Hamiltonian (\ref{H_pert_gen}) is restricted to the truncated state space. 
For a primary field $\psi(0,0)$ of conformal dimension $\Delta^{\rm UV}$ the matrix elements between states of 
the form (\ref{desc_states}) can be calculated using  three point functions and standard commutation relations
\be
[L_{n} - L_{0}, \psi(0,0)] = \Delta^{\rm UV} n \psi(0,0) \, .
\ee
The resulting finite matrix is then diagonalised numerically. 

The main TCSA observables are energy levels $e_{i}$ of the dimensionless Hamiltonian $L H/\pi$. 
These levels are functions of the dimensionless coupling 
\be
\lambda=\mu L^{1-\Delta^{\rm UV}}
\ee
   The ground state energy $e_{0}$ is divergent in the continuum theory when $\Delta^{\rm UV}\ge 1/2$ and if we work with raw TCSA 
   data, we focus on the energy gaps $e_{i} - e_{0}$ that are independent of this divergence. 
   The values of the gaps $e_{i}-e_{0}$  interpolate between the (approximations to) 
   scaling dimensions of the UV and IR fixed points. While in the continuum theory the IR fixed point is located 
   at an infinite value of the coupling, in TCSA we get the closest approach to the IR BCFT at some finite value which 
   depends on $\Delta_{\rm max}$. The main signature of this approach is the near crossings of the gaps $e_{i}-e_{0}$ 
   that can be matched  with the Virasoro multiplicities in the IR BCFT.  
   




 TCSA has been applied to a variety of models to obtain non-perturbative information of various  types matching with good accuracy to analytic results when available.  We refer the reader to \cite{TCSA_review} for a review of the method and results obtained.  While TCSA has proven to be largely successful, it would be fair to say there is no 
good understanding of why the method works so well. The question of error estimates related to the finite truncation level has been addressed in a number  of papers (see \cite{Gerard2}, \cite{GiokasWatts}, 
\cite{HRvR}, \cite{RV}, \cite{NLO} and references therein). Those papers focus on the regime where 
the ratio of the characteristic energy scale set by the coupling to the truncation level is small and a perturbative expansion in this ratio is possible. While a good theoretical understanding has been 
achieved here, that work does not shed much light on a very high coupling regime, in particular, where the system approaches an IR point. In the present paper we focus on the  TCSA behaviour for very large couplings where analytic control of truncation errors is currently out of reach. Our results are thus mostly numeric but we do attempt to get theoretical   
understanding of the patterns emerging  at least at the level of phenomenology.

In the next section we discuss some generalities of the large coupling asymptotics, setting out our theoretical expectations before discussing results for specific models.
In section \ref{Approach_section} we focus on  the TCSA description of the approach to an IR fixed point. 
We investigate  the infrared exponents related to the leading irrelevant operator along which the 
system enters the fixed point. We describe some general methods of extracting these exponents from 
numerical data and test them for the examples of boundary flows in the Ising and tricritical Ising models. 
 
In TCSA the closest approach to an IR fixed point emerges at a finite value of the coupling. Increasing the coupling past that value, in the case of boundary flows, it has been observed \cite{Gerard2} that the low lying spectrum rearranges itself to approximate that of another fixed point. For some perturbations such ``flows beyond'' pass through a sequence of several fixed points. We investigate this mysterious phenomenon in section \ref{FlowsBeyond_section} drawing mostly on numerical results for boundary flows in the Ising and Tricritical Ising models. In section \ref{exact_beyond} however we present an exactly solvable model of a truncated theory that exhibits flows beyond the fixed point. We end the paper with some concluding remarks in section \ref{conclusions_sec}.


\section{General considerations}\label{beyond_general}
\setcounter{equation}{0}
In this section we discuss some generalities of the  large coupling behaviour in perturbed BCFTs. Our considerations 
here are mostly analytic, though they include a dose of  heuristics inspired by numerical data. 

In a continuum renormalised theory, if we start out in the UV by perturbing a boundary CFT by a boundary field of 
dimension $\Delta^{\rm UV}$ the corresponding coupling $\mu$ of dimension $(\mbox{mass})^{1-\Delta^{\rm UV}}$ sets an energy scale 
\be
E_{\mu} = (\mu)^{\frac{1}{1-\Delta^{\rm UV}}} \, .
\ee
Near the infrared fixed point described by a BCFT with the Virasoro dilation operator $L_{0}^{\rm IR}$ we expect the 
theory to be described by an effective Hamiltonian of the form 
\be\label{hIRgen}
h^{\rm IR}= L_{0}^{\rm IR} + C_{1}\mu^{-t_1}L\phi_{1}^{\rm IR}(0) + C_{2}\mu^{-t_2}L\phi_{2}^{\rm IR}(0) + \dots 
\ee
where $\phi_{i}^{\rm IR}$ stand for irrelevant operators with scaling dimensions $\Delta^{\rm IR}_{1}<\Delta^{\rm IR}_{2} < \dots$ 
inserted at $\sigma=0, \tau=0$. The operator $\phi_{1}^{\rm IR}$ is the leading irrelevant operator. Matching the dimensions  implies 
\be\label{ti}
t_{i} = \frac{\Delta_{i}^{\rm IR}-1}{1-\Delta^{\rm UV}} \, .
\ee
We can also rewrite (\ref{hIRgen}) in terms of the dimensionless coupling:
\be\label{hIRexp}
h^{\rm IR}= L_{0}^{\rm IR} + C_{1}\lambda^{-t_1}V_{1}^{\rm IR} + C_{2}\lambda^{-t_2}V_{2}^{\rm IR} + \dots 
\ee
where $V_{i}^{\rm IR}=\phi_{i}(0)L^{\Delta_{i}^{\rm IR}}$ are dimensionless operators.  
It should be said that in the  subleading terms starting with $V_{2}$ the simple power functions $\lambda^{-t_i}$ can be decorated by logarithms (see e.g. \cite{Feverati_etal_IR}) 
but we do not expect this to happen at the leading irrelevant term. 

It follows from (\ref{hIRexp}), using perturbation theory, that for large $\lambda$ the energy gaps should behave as
\be \label{energy_approach}
e_{i}-e_{0} = \Delta_{i}^{\rm IR} + A_{i}\lambda^{-t_1} + \dots
\ee
where $A_{i}$ are some constants and the ellipsis stands for subleading terms suppressed by larger inverse powers of $\lambda$.


Consider now the same  perturbed BCFT regulated by level truncation. 
Numerical results show that the energy levels  approach the dimensions of the IR fixed point in a neighbourhood 
of some finite reference value of the coupling $\lambda^{*}$. This value depends both on the truncation parameter 
$\Delta_{\rm max}$ and on the energy level. We assume that we focus on some particular band of low lying energies 
which all  approach the dimensions of the IR fixed point  in a neighbourhood of  $\lambda^{*}$.
We expect truncation errors to modify (\ref{energy_approach}). Heuristically we can assume that (\ref{energy_approach}) 
is replaced by 
\be \label{energy_approach2}
e_{i} - e_{0} = \Delta_{i}^{\rm IR}+  (\lambda - \lambda_{*})A_{i}^{(-1)} +  A_{i}\lambda^{-t_1} + \dots
\ee
Here the first term is small when $\lambda$ is near $\lambda_{*}$ but is linearly growing when $\lambda$ deviates from that value. 
The linearity of this correction is motivated by the fact that in the truncated Hamiltonian both the unperturbed Hamiltonian and the 
perturbation are finite matrices and for large $\lambda$ the perturbed eigenvalues become approximately linear in $\lambda$. 
The numerical coefficients $A_{i}^{(-1)}$ and $A_{i}$ can be expected to depend on $\Delta_{\rm max}$. In particular     $A_{i}^{(-1)}$ 
should be suppressed by inverse powers of  $\Delta_{\rm max}$. 

 To describe the above corrections at the Hamiltonian level  we  assume that the following relation holds 
for matrices on the truncated Hilbert space
\be \label{hUVIR}
h=L_{0}^{\rm UV} + \lambda V^{UV} = e^{T(\lambda)}(L_{0}^{\rm IR} + f_{1}(\lambda)V_{1}^{\rm IR} +  f_{2}(\lambda)V_{2}^{\rm IR} + \dots) 
e^{-T(\lambda)}
\ee
where, as in (\ref{hIRexp}), the matrices $V_{i}$ come from local irrelevant operators of the IR fixed point and $T(\lambda)$ is an anti-hermitian matrix so that $e^{T(\lambda)}$ is a unitary operator (for each $\lambda$).  Furthermore  the  functions $f_{i}(\lambda)$ are expected to be  small whenever $|\lambda-\lambda^{*}|<\rho$ for some $\rho$.  It should be stressed that we consider only a number of low energy eigenvalues and treat the representation in 
 (\ref{hUVIR}) above only as 
an effective Hamiltonian for the low lying energy levels. 

  Assuming that in the $\Delta_{\rm max} \to \infty$ limit we recover the field theory expansion 
(\ref{hIRexp}) we can write 
\be \label{fii}
f_{i}(\lambda) = \tilde f_{i}(\lambda-\lambda^{*}) + C_{i} \lambda^{-t_{i}}
\ee
where the functions $\tilde f_{i}(\lambda-\lambda^{*})$ are small for  $|\lambda-\lambda^{*}|<\rho$ and are suppressed by inverse powers of $\Delta_{\rm max}$. 
In this picture whenever $f_{i}$ are small enough the eigenvalues of $h$ can be  approximated by first order perturbation theory around the 
eigenvalues of $L_{0}^{\rm IR}$ and thus will have approximately the same functional form as (\ref{fii}). In particular they may have 
a linearly growing term as in (\ref{energy_approach2}). If we are outside the infrared fixed point dominance region, given by $|\lambda-\lambda^{*}|<\rho$, 
perturbation theory will no longer work and the energy curves $e_{i}(\lambda)$ may significantly deviate from the IR fixed point values. 
If the function $\tilde{f}_{1}(\lambda-\lambda^{*})$ is positive and growing while $C_{i}<0$ 
then\footnote{$C_{i}<0$ is the case when the energies approach 
the fixed point values from below as in the Ising and tricritical Ising boundary flows to be discussed later in the next section.}
 for some $\lambda=\lambda^{**}$ the function
$f_{1}$ will vanish and then change sign. If near $\lambda^{**}$ the leading irrelevant coupling  $f_{1}$ changes faster than the subleading ones 
and the latter remain small then this regime can be {\it perturbatively} described as a reflected RG flow where we perturb the IR fixed point by $V_1$ with 
the opposite sign of the coupling which grows in magnitude with $\lambda$. 

We have argued above that the linearity of the leading truncation correction in (\ref{energy_approach2}) could be explained from 
the asymptotic linearity of the eigenvalues at large $\lambda$. The slopes $A_{i}^{(-1)}$ would then be given by the differences of 
the corresponding eigenvalues of the interaction matrix $V^{\rm UV}$. This is certainly true for very large values of $\lambda$.
 However, in numerical data to be presented later in the paper, we observe several approximately linear regimes past 
the vicinity of the physical IR fixed point at $\lambda=\lambda^{*}$, before the linear regime described by the domination of $V^{\rm UV}$ is reached. 
Each of these approximately linear regimes has its own set of slopes. We will discuss these regimes, related to what we call flows beyond, in detail, 
in section \ref{FlowsBeyond_section}.  
Here we would like  to refine our simple explanation for the linearity, using  relation (\ref{hUVIR}) between  the effective Hamiltonians. 
We will argue that, under certain additional assumptions  to be spelled out shortly,  the functions $e_{i}(\lambda)$ are approximately linear near $\lambda^{**}$. 
To that end differentiate both sides of (\ref{hUVIR}) 
and set $\lambda=\lambda^{**}$. This gives us 
\be
V^{\rm UV} = e^{T(\lambda^{**})}( [T'(\lambda^{**}),L_{0}^{\rm IR} + f_{1}(\lambda^{**})V_{1}^{\rm IR} + \dots ]  + f_{1}'(\lambda^{**})V_{1}^{\rm IR} 
+ \dots ) e^{-T(\lambda^{**})}
\ee 
where ellipses stand for contributions of subleading irrelevant operators. Substituting this back into the left hand side of (\ref{hUVIR}) 
we rewrite the Hamiltonian as 
\bea \label{hUVIR2}
&& h=L_{0}^{\rm UV} + \lambda V^{UV} = (L_{0}^{\rm UV} + \lambda^{**} V^{UV}) + (\lambda-\lambda^{**})V^{\rm UV} \nonumber \\
&& = e^{T(\lambda^{**})}( L_{0}^{IR} +      g_{1}(\lambda)V_{1}^{\rm IR}   + g_{2}(\lambda)V_{2}^{\rm IR} + \dots  \nonumber  \\
&& +(\lambda-\lambda^{**})[T'(\lambda^{**}),L_{0}^{\rm IR} + f_{1}(\lambda^{**})V_{1}^{\rm IR} + f_{2}(\lambda^{**})V_{2}^{\rm IR} \dots ]  
 ) e^{-T(\lambda^{**})}
\eea
where 
\be \label{gii}
g_{1}(\lambda) =  (\lambda-\lambda^{**})f_{1}'(\lambda^{**}) \, , \quad 
g_{i}(\lambda) = f_{i}(\lambda^{**}) +  (\lambda-\lambda^{**})f_{i}'(\lambda^{**})\, , \enspace i>1 \, .
\ee
Here we used the assumption that $f_{1}(\lambda^{**})=0$.

If the generator of unitary rotations, $T(\lambda)$, can be expanded in local operators then the commutator term can also be expanded in 
local operators and we get a local expansion in operators\footnote{Also we should mention that there is a possibility of switching on a relevant operator with a coupling suppressed by 
$\Delta_{\rm max}$. For the $\psi_{1,3}$ flows we consider in the tricritical and in the critical Ising models the IR fixed points do not have 
any relevant operators so, for simplicity, in our general discussion we assume they are absent. } $V_{i}^{\rm IR}$ with modified functions 
$g_{i}(\lambda)$ that are again 
linear in $\lambda$.  Alternatively the commutator term may be negligibly small in the vicinity of $\lambda^{**}$ if the functions $g_{i}(\lambda)$ 
change more rapidly than $T(\lambda)$. We will assume that one of these two scenarios is realised and we can neglect the 
commutator terms. This implies that we have linear trajectories  in the space of IR Hamiltonians parameterised by the IR couplings $g_{i}$ given by (\ref{gii}) 
where $g_{1}$ vanishes at $\lambda^{**}$. 

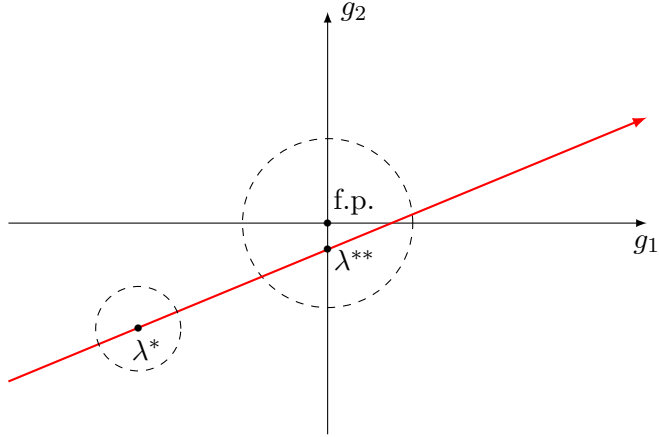
\begin{figure}[H] 
 \begin{center}
\begin{tikzpicture}[>=latex,scale=1.4]
Middle vertical nodes
\draw[dashed] (0,0) circle [radius=0.8] ;
\draw (0.25, 0.2) node {f.p.};
\draw (0,0) node {\tiny \textbullet} ;
\draw[->] (-3,0)  --   (3,0);
\draw (3,-0.2) node {$g_1$};
\draw (0.25,2) node {$g_2$};
\draw[->] (0,-2)  --   (0,2);
\draw[->, red, thick] (-3, -1.5) -- (3, 1);
\draw (0,-0.25) node {\tiny \textbullet};
\draw (0.25, -0.35) node {$\lambda^{**}$};
\draw (-1.78,-1) node {\tiny \textbullet};
\draw[dashed] (-1.78,-1) circle [radius=0.4];
\draw (-1.7, -1.2) node {$\lambda^{*}$};
\end{tikzpicture}
\caption{A sample TCSA trajectory in $\{g_{i}\}$-space. Two perturbative regimes are marked by circles}
\label{RGflow_picture}
\end{center}
\end{figure}

Assuming that near $\lambda^{**}$ all irrelevant couplings $g_{i}$ are small with $g_1$ still dominating we have the second perturbative regime 
described by a reflected flow with $g_1$ of the opposite sign to that of the physical approach to the fixed point, and changing linearly with $\lambda$.
In that region the energy gaps $e_{i}-e_{0}$ are also changing linearly with the coupling\footnote{Unless there is some symmetry 
that forces the diagonal matrix elements of $V_1$ to vanish in which case the perturbative expansion starts with some power $\lambda^n$ with  $n>1$.}.
We offer this scenario as a tentative explanation of  the approximately linear regime of $e_{i}-e_{0}$  
that we find in the numerical results immediately past the point of closest approach to the IR fixed point. The above discussion is illustrated
on Fig. \ref{RGflow_picture}. It should be noted in regard to this picture that although in the space of $g_{i}$ couplings  the point $\lambda^{*}$ 
appears to be far from the obvious fixed point $g_{i}=0$, due to the action of unitary transformations it is actually close to it. 


 

It is  important to note that the effective IR Hamiltonians appearing on the right hand sides of (\ref{hUVIR}) and (\ref{hUVIR2}) are finite 
matrices defined in the truncated Hilbert space of the UV fixed point. 
At the IR fixed point this truncation is representable by some {\it different}, complicated truncation prescription. Thus if we were to write (\ref{hUVIR}) or (\ref{hUVIR2}) in 
the continuum theory representing the IR CFT,  each operator $\phi_{i}^{\rm IR}$ would be represented in some complicated truncation regularisation. If the latter still respects locality  it can be modelled in the continuum theory by writing out an expansion   
of the regulated operators in a series of continuum irrelevant operators with coefficients suppressed by $\Delta_{\rm max}$.

   
 The above arguments along with the numerical data to be considered in the forthcoming sections prompt us to formulate a working hypothesis: 
 the flows beyond are generated by the leading irrelevant perturbation, $V_{1}^{\rm IR}$, along which the original RG flow 
 approaches the IR fixed point, with the opposite sign of the coupling.  This coupling grows linearly with the original coupling $\lambda$ in a neighbourhood of some value $\lambda=\lambda^{**}$. 
 Here we also assume that the higher dimension irrelevant operators can be neglected and that some particular truncation scheme is used 
 at  the infrared fixed point to model the flows beyond.  We will test this working hypothesis in sections  \ref{beyond3} - \ref{beyond5}. 
 

\section{Approach to IR fixed point} \label{Approach_section}
\setcounter{equation}{0}

\subsection{Ising model with a boundary magnetic field}\label{Ising1_sec}
A particularly simple example amenable to analytic control is the critical Ising model with  a boundary 
magnetic field \cite{GZ, CZ, Chat, me1, Toth, Toth2, AKB}. The critical Ising model in the bulk is described by free massless Majorana fermions
$\psi$, $\bar \psi$. On the upper  half plane $\{ (x,y)| y\ge 0\}$ with complex coordinate $z=x+iy$ 
the model admits two conformal boundary conditions: $\psi(z)=\bar \psi(\bar z)$, $z=\bar z$ that
corresponds to having the boundary spin fluctuating freely, and $\psi(z)=-\bar \psi(\bar z)$, $z=\bar z$
that describes keeping the boundary spin fixed (up or down). 
We will use  the notation: $(+)$, $(-)$, $(f)$ for the spin up, spin down and free boundary condition respectively. 
 
To describe the doubly degenerate vacuum on 
the half plane we introduce, following \cite{GZ}, \cite{CZ},  a boundary fermion $a(x)$ with a two point function
\be
\langle a(x)a(x')\rangle = {\rm sign}(x-x') \, .
\ee
The boundary spin  operator $\sigma_{\rm B}(x)$ is then given by 
\be
\sigma_{\rm B}(x) = ia(x)(\psi(x) + \bar \psi(x)) \, .
\ee
In boundary CFT language this  is a primary boundary field of dimension $1/2$ that lives on the free boundary condition. 
We can perturb the latter by this field with a coupling $h$ that gives the value of a boundary magnetic field. 
The resulting Lagrangian is 
\begin{equation}
S = \frac{1}{2\pi}\int\limits_{-\infty}^{\infty}\!\!dx\int\limits_{0}^{\infty}\!\!dy \  [\psi\bar{\partial}\psi + \bar{\psi}\partial\bar{\psi}] + 
\int\limits_{-\infty}^{\infty}\!\!dx \Bigl[  -\frac{i}{4\pi}\psi\bar{\psi}+\frac{1}{2}a\dot{a} + ih  a(\psi+\bar{\psi)} \Bigr]\, .
\end{equation}

To describe this model in TCSA we put it on an infinite strip $\{ (\sigma, \tau)| 0\le \sigma \le L\}$ of width $L$. 
It is related to the upper half plane by a conformal mapping $w=(L/\pi){\rm ln} \, z$ where $w=\tau + i\sigma$. 
We put the free boundary condition on the lower end of the strip: $\sigma=0$. This is the end we perturb by the 
boundary magnetic field. For the spectator boundary condition at the other end, at $\sigma=L$, we can 
choose either the free or fixed boundary condition.  

 In Hamiltonian quantisation with euclidean time $\tau$, a
 free spectator corresponds to having Neveu-Schwarz (NS) 
fermions with mode expansions 
\bea
&& \psi(w) = \sqrt{\frac{\pi}{L}}
\sum_{k=0}^{\infty} \Bigl[  e^{-\frac{(k+1/2)\pi}{L}(\tau + i\sigma)} a_{k+1/2}  +  
e^{\frac{(k+1/2)\pi}{L}(\tau + i\sigma)} a_{k+1/2}^{\dagger}   \Bigr] \, , \nonumber \\
&& \bar \psi(w) = \sqrt{\frac{\pi}{L}} \sum_{k=0}^{\infty} \Bigl[ e^{-\frac{(k+1/2)\pi}{L}(\tau - i\sigma)} a_{k+1/2}  + 
e^{\frac{(k+1/2)\pi}{L}(\tau - i\sigma)} a^{\dagger}_{k+1/2}  \Bigr]\, 
\eea
where $a_{k+1/2}$, $a^{\dagger}_{k+1/2}$ satisfy canonical anti-commutation relations. The boundary fermion field $a(\tau)$ gives rise
to a single fermionic mode $a$ satisfying $a^2=1$. The corresponding Hamiltonian reads 
\begin{equation}\label{NS Ham}
H^{\rm NS} = \frac{\pi}{L}\Bigl[ \sum_{k=0}^{\infty} (k+1/2) a_{k+1/2}^{\dagger}a_{k+1/2} - \frac{1}{48}
+ i\alpha \sum_{k=0}^{\infty} (a_{k+1/2}^{\dagger} + a_{k+1/2})a \Bigr]
\end{equation}
where\footnote{Note that $h$ and $\alpha$ in this paper have the opposite sign to those in \cite{AKB}} 
\be
\alpha = h\sqrt{\frac{\pi}{L}}
\ee
is a dimensionless coupling. The Hamiltonian (\ref{NS Ham}) is defined on a  physical subspace of the Fock space spanned by the basis vectors 
\bea\label{NSbasis}
&& a^{\dagger}_{k_1+1/2}a^{\dagger}_{k_2+1/2}\dots a^{\dagger}_{k_{N}+1/2}|0\rangle \quad N-\mbox{ even}, \enspace 
k_{1}>k_2> \dots > k_{N} \, , \nonumber \\
&& a^{\dagger}_{k_1+1/2}a^{\dagger}_{k_2+1/2}\dots a^{\dagger}_{k_{N}+1/2}|a\rangle \quad N-\mbox{ odd}, \enspace 
k_{1}>k_2> \dots > k_{N}
\eea
where $|0\rangle$ is the Fock space vacuum and $|a\rangle = a|0\rangle$. The physical space contains two irreducible 
representations of the Virasoro algebra: the identity tower spanned by the states with  even numbers of oscillators and the 
$\epsilon$-tower spanned by the states with odd numbers of oscillators.

Similarly, choosing a fixed spin spectator at $\sigma=L$ gives rise to Ramond fermions with mode expansions 
 \bea
&& \psi(w) = \sqrt{\frac{\pi}{L}}
\sum_{n=1}^{\infty} \Bigl[  e^{-\frac{n\pi}{L}(\tau + i\sigma)} b_{n}  +  
e^{\frac{n\pi}{L}(\tau + i\sigma)} b_{n}^{\dagger}   + b_{0}\Bigr] \, , \nonumber \\
&& \bar \psi(w) = \sqrt{\frac{\pi}{L}} \sum_{n=1}^{\infty} \Bigl[ e^{-\frac{n\pi}{L}(\tau - i\sigma)} b_{n}  + 
e^{\frac{n\pi}{L}(\tau - i\sigma)} b^{\dagger}_{n} + b_{0}  \Bigr]
\eea
where $b_{n}$, $b_{n}^{\dagger}$ satisfy the canonical anti-commutation relations with the zero mode normalised so that $b_{0}^2 = 1/2$. 
The Hamiltonian in this case is 
\begin{equation}\label{R Ham}
H^{\rm R} = \frac{\pi}{L}\Bigl( \sum_{n=1}^{\infty} nb_n^{\dagger}b_n + \frac{1}{24}  + i\alpha [\sum_{n=1}^{\infty} (b_n^{\dagger} + b_n) + b_0]a \Bigr)
\end{equation}
and the basis in the physical subspace of the Fock space can be chosen as 
\bea
&& b^{\dagger}_{n_1}b^{\dagger}_{n_2}\dots b^{\dagger}_{n_M}|\sigma \rangle \quad M-\mbox{ even}, \enspace 
n_{1}>n_2> \dots > n_{M}>0 \, , \nonumber \\
&& b^{\dagger}_{n_1}b^{\dagger}_{n_2}\dots b^{\dagger}_{n_M}|\mu \rangle \quad M-\mbox{ odd}, \enspace 
n_{1}>n_2> \dots > n_{M}>0
\eea
where $|\sigma\rangle$ is the Fock state vacuum  and $|\mu\rangle = -ia|\sigma\rangle$. This space furnishes a single Virasoro 
tower of the  primary state $|\sigma\rangle$ with weight $1/16$.
The zero mode acts on the vacuum as 
\be
b_{0} |\sigma\rangle = \frac{1}{\sqrt{2}}|\mu\rangle \, .
\ee

For each choice of spectator boundary condition the Hamiltonian can be diagonalised by a Bogolyubov transformation 
\cite{Toth}, \cite{AKB}. The diagonalising modes $b^{\dagger}_{\alpha, i}$ carry energies $\omega_{i}$ 
that are non-negative solutions to a transcendental equation that for the free spectator reads as 
\be\label{NSspec}
\tan(\pi \omega) = -\frac{\omega}{2\pi\alpha^2} 
\ee
and for a fixed spectator as
\be\label{Rspec} 
\tan(\pi\omega) = \frac{2\pi\alpha^2}{\omega} \, . 
\ee

For $\alpha \to \pm \infty$ solutions to (\ref{NSspec}) interpolate between half integers at $\alpha=0$ and integers at $\alpha = \pm \infty$. 
The energy eigenstates are obtained by acting on the perturbed vacuum $|0\rangle_{\alpha}$ by an even number of raising operators 
$b^{\dagger}_{i}$. At the endpoints $\alpha=\pm \infty$ we thus have the usual physical space of Ramond fermions and the model describes 
a flow from free to fixed boundary condition in the far infrared. It is completely symmetric under changing the sign of $\alpha$. The latter 
specifies  the infrared boundary condition to be   spin up or down. The boundary fermion $a$ is absorbed into the zero mode $b^{\dagger}_{0}$. 

For the fixed spectator the spectral equation (\ref{Rspec}) interpolates between integer solutions at $\alpha=0$ and the half-integer ones 
at $\alpha = \pm \infty$. The physical spectrum however depends on the sign of $\alpha$. For negative $\alpha$  the Fock space 
vacuum $|0\rangle_{\alpha}$ with respect to the diagonalising oscillators does not belong to the physical space and thus the physical vacuum is 
$b^{\dagger}_{\alpha,1}|0\rangle_{\alpha}$ where $b^{\dagger}_{\alpha, 1}$ is the creation operator for the lowest excitation energy $\omega_1$. 
The physical space for $\alpha \to -\infty$ contains a single Virasoro tower of the $\epsilon$ representation and describes the state space with the
opposite direction fixed boundary conditions on the two ends of the strip. 
For positive $\alpha$  the physical vacuum coincides with the Fock space vacuum and in the limit $\alpha \to \infty$ we 
obtain the Virasoro tower of the identity field that describes the fixed spin boundary condition with the same direction on the two ends. 

It was shown in \cite{Toth} that the perturbed boundary condition approaches the infrared fixed point along the leading irrelevant 
operator, the stress energy tensor $T$. More precisely the leading effective Hamiltonian near $\alpha = \infty$ is 
\be \label{Ising_h_eff}
h_{\rm eff}^{\rm lead} = \frac{1}{24} + \sum_{k=1}^{\infty}k\, b^{\dagger}_{k}b_{k} 
- \frac{g}{2}:\!\left(\left[ \sum_{k=1}^{\infty} (b^{\dagger}_{k} +b_{k}) + b_{0}\right] \sum_{l=1}^{\infty} l(b^{\dagger}_{l} - b_{l})\right)\!:
\ee
with 
\be
g=-\frac{1}{2\pi^2\alpha^2} \, . 
\ee
The subleading terms in the effective Hamiltonian are discussed in \cite{AKB}. The leading exponent shows up in the 
asymptotic behaviour of the perturbed energy levels. From (\ref{Rspec}) we obtain a large $\alpha$ expansion for the fixed spectator 
\begin{equation}\label{expansion}
\omega _k = (k-1/2)(1-\frac{1}{2\pi \alpha ^2}+\frac{1}{(2\pi \alpha ^2)^2}+\frac{1}{(2\pi \alpha ^2)^3}[\frac{\pi ^2}{3}(k-1/2)^2 -1]+\dots)
\end{equation}
where the missing terms are of the order $1/\alpha^8$. A similar expansion can be worked out for the excitation energies in the case of 
a free spectator with the leading correction to integer values being of the order of $1/\alpha^2$. 

\subsection{TCSA in the Ising model. Approach to fixed point. }
For the Ising model with a boundary magnetic field, for the case of a fixed spectator (Ramond fermions) there is only one primary tower whereas for the case of the free spectator described by NS-fermions we have two primary fields.  The dimensions of the truncated state space for the free and fixed spectators are displayed in the Appendix \ref{App1}.
Later we will also use modified truncation schemes of the type (\ref{modif_scheme}) labelled by 
$N_{\rm max}^{1}$ and $N_{\rm max}^{\epsilon}$ 
-- the bounds for descendant levels in the identity and the energy density towers. 

\begin{figure}[H]
  \begin{subfigure}[b]{0.47\textwidth}
    \includegraphics[width=\textwidth]{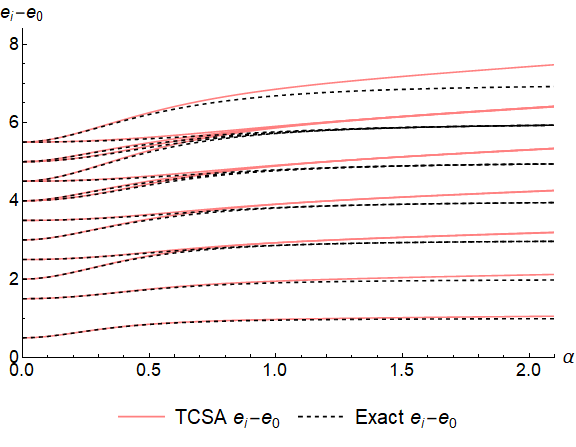}
    \caption{NS Sector (free spectator).  $e_i-e_0$ versus $\alpha$ for $\Delta_{\rm max}=38$}
  \end{subfigure}
  \qquad	
  \begin{subfigure}[b]{0.47\textwidth}
    \includegraphics[width=\textwidth]{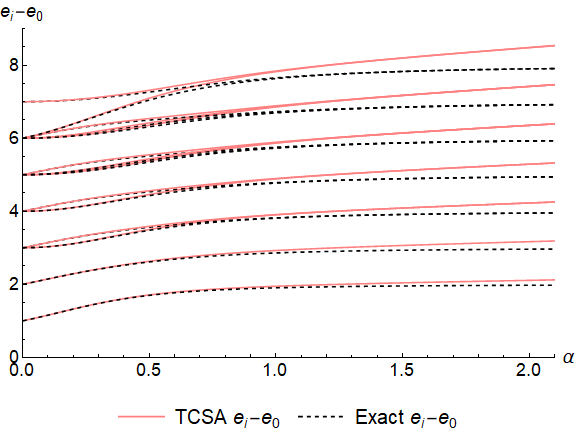}
    \caption{Ramond Sector (fixed spectator).  $e_i-e_0$ versus $\alpha$ for $\Delta_{\rm max}=40$}
   \end{subfigure}
  \caption{}
   \label{NSRplots}
\end{figure}

In this section we concentrate 
on the approach to the fixed point in the boundary magnetic field model. 
Below, on Fig. \ref{NSRplots}  ,  are  the plots of the dimensionless energy gaps $e_{i} - e_{0}$ for the first 14 excited  energy levels.

On these plots we also put the exact solutions (dashed lines). In the first plot we observe that the TCSA energy gaps remain close to the exact solutions 
until the onset of the asymptotic regime corresponding to the IR fixed point which is marked by levelling of the energy curves and 
by the change in  the multiplicities of 
the energy levels. Although we do not get the exact degeneration (that in the exact solution also happens only asymptotically at infinite coupling)  we  clearly see
the levels nearly merging giving the degeneracies of the expected IR fixed point described by Ramond fermions.
However, while the levels  initially converge, as we increase the coupling they start diverging from each other and at the same time  the TCSA curves start deviating from the exact 
solutions. The latter have horizontal asymptotes, with the corrections following  the  expansion (\ref{expansion}), while the TCSA gaps 
become, to a good approximation, linear with the coupling with positive slopes. 
The higher energy level we take the smaller are the values of $\alpha$ at which 
 deviation from the exact solution starts.  Similar behaviour is observed in the second plot.   Focussing on a few low energy levels we can approximately determine a scale $\alpha_{*}$ that is the largest scale 
around  which these energy levels are the closest to the IR fixed point exact values. This value depends on $\Delta_{\rm max}$. For a fixed number of 
lowest levels, the higher $\Delta_{\rm max}$ is, the higher  the value of $\alpha_{*}$ can be chosen. 
As explained in section \ref{beyond_general}
we expect the 
energy levels to behave approximately as   
\be \label{fp_ansatz}
e_{i} - e_{0} \approx  \Delta_{i}^{\rm IR}+ (\alpha - \alpha_{*}) A_{i}^{(-1)} +  \frac{A_{i}^{(2)}}{\alpha^t}  + \dots  
\ee
where the continuum theory prediction  based on (\ref{Ising_h_eff}) is $t=2$. In principle one may expect a more general modification of the form 
\be \label{fp_ansatz1}
e_{i} - e_{0} \approx  \Delta_{i}^{\rm IR}+ (\alpha - \alpha_{*}) A_{i}^{(-1)} + \frac{A_{i}^{(1)}}{\alpha} + \frac{A_{i}^{(2)}}{\alpha^2}  + \dots  
\ee
where the coefficients $A_{i}^{(-1)}$, $A^{(1)}_{i}$, $A^{(2)}_{i}$ depend on $\Delta_{\rm max}$ and the ellipses stand for the terms suppressed by higher powers of 
$\alpha$. To reproduce the continuum results the coefficients $A_{i}^{(-1)}$, $A^{(1)}_{i}$  can be both be expected to be suppressed by powers of 
$\Delta_{\rm max}$ while $A^{(2)}_{i}$ should be constant up to terms suppressed by powers of $\Delta_{\rm max}$.
It is quite hard to extract next to leading order power corrections from numerical data so we are going to test the simpler ansatz (\ref{fp_ansatz}). 
 If we subtract the linear part in some way then we can estimate $t$ from the numerical data. 

\begin{figure}[H]
  \begin{subfigure}[b]{0.47\textwidth}
    \includegraphics[width=\textwidth]{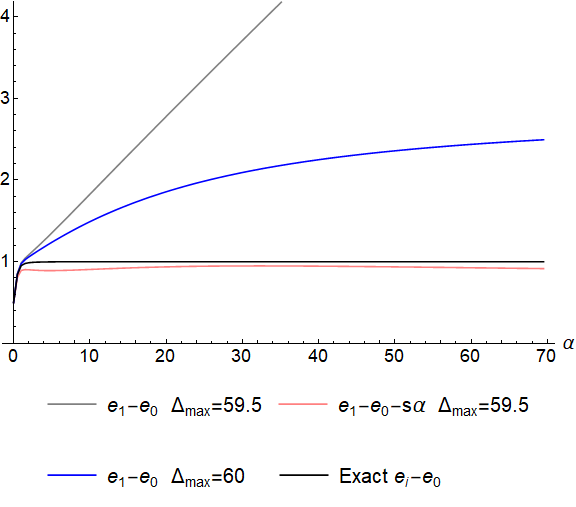}
    \caption{$e_{1}-e_{0} - s\alpha$ vs $\alpha>0$ with $s$ given by  the interaction matrix eigenvalues for the free spectator for $\Delta_{\rm max}=59.5$ and $\Delta_{\rm max}=60$}
  \end{subfigure}
  \qquad	
  \begin{subfigure}[b]{0.47\textwidth}
     \includegraphics[width=\textwidth]{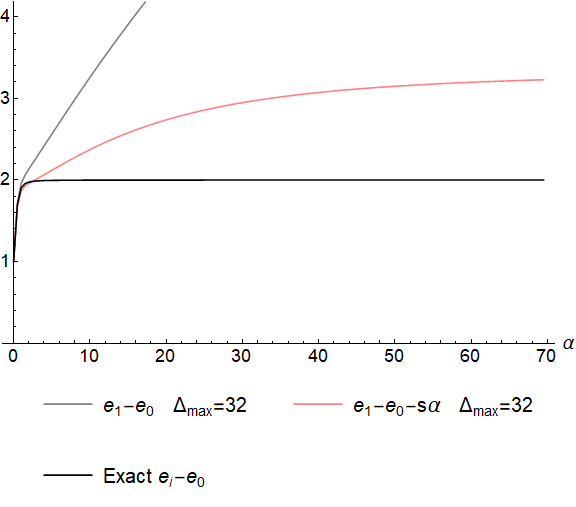}
    \caption{$e_{1}-e_{0} - s\alpha$ vs $\alpha>0$ with $s$ given by  the interaction matrix eigenvalues for the fixed spectator with $\Delta_{\rm max}=32$}
  \end{subfigure}
  \caption{}
  \label{asympt_plots}
\end{figure}

As mentioned before, the linear part in (\ref{fp_ansatz}), which ultimately takes the energy levels away from the fixed point, could be related to the domination of the truncated interaction 
matrix $V$. Since in TCSA we are dealing with a finite matrix of the form $h=L_{0} + \alpha V$, when $\alpha$ is large enough the interaction matrix $V$ dominates 
and the eigenvalues are well approximated by $\alpha$ times the eigenvalues of $V$. So that the  $\alpha\to \pm \infty$ behaviour is definitely linear. 
To test this idea we plot on Fig. \ref{asympt_plots} the quantity $e_1-e_{0}-s\alpha$ where $s$ is the difference of the two lowest eigenvalues of $V$.  We note that the two lowest eigenvalues for the free spectator with integer $\Delta_{\rm max}$ are degenerate and hence, we plot only the unmodified spectrum.

The plots show that for larger values of $\alpha$ the energy gap flattens and the interaction matrix dominates as expected.  
Zooming in to the region near the fixed point, we observe that there is a residual difference between $s$ and the slope of $e_1-e_0$ near the fixed point. 
There are thus {\it two different regimes} in which the energy gaps are approximately linear with the first regime being near the appearance of 
the physical IR fixed point while 
the second regime where eigenvalues of $V$ dominate can be related to the flow beyond which we are going to discuss in detail later.

To estimate $t$ from numerical data we use  two slightly different procedures. 
In the first method we calculate numerically 
\be
{\cal D}^{1}_{s}(e_{1}-e_{0}) \equiv \alpha\frac{d}{d\alpha}{\rm ln}\left( \alpha\frac{d}{d\alpha}[e_{1}(\alpha) -e_{0}(\alpha)  - s\alpha]\right)
\ee
where $e_{i}(\alpha)$ are the dimensionless TCSA  energy levels and $s$ is a numerically obtained slope of the function 
$e_{1}(\alpha) - e_{0}(\alpha)$ past $\alpha=\alpha_{*}$. As we do not have a good understanding of the linearity in $\alpha$ 
of the leading truncation correction  we are going to use two different prescriptions for $s$: taking the difference of 
the two lowest eigenvalues of the interaction matrix and using a regression analysis for  TCSA numerics near the appearance of the IR fixed point.

In the second method we subtract the linear part by applying an additional differential operator $1-\alpha d/d\alpha $:
\be\label{d2}
{\cal D}^{2}(e_{1}-e_{0}) \equiv \alpha\frac{d}{d\alpha}{\rm ln}\left( \alpha\frac{d}{d\alpha}(1-\alpha\frac{d}{d\alpha})[e_{1}(\alpha) -e_{0}(\alpha) ]\right)
\, . 
\ee
These differential operators are chosen so that   substituting into them a function of the form (\ref{fp_ansatz}), we obtain $-t$. In the presence 
of subleading terms we expect this to be approximately true so that the function changes slowly near the value $-t$.

\begin{figure}[H]
\centering
\includegraphics[width=0.85\textwidth]
{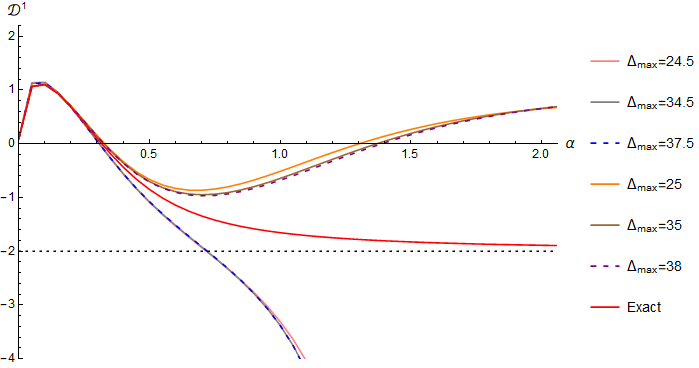}
\caption{\label{fig:your-figure4}${\cal D}^{1}_{s}(e_{1}-e_{0})$ versus $\alpha>0$ for the free spectator. Linear term removed using interaction matrix eigenvalues.}
\end{figure}

\begin{figure}[H]
\centering
\includegraphics[width=0.85\textwidth]
{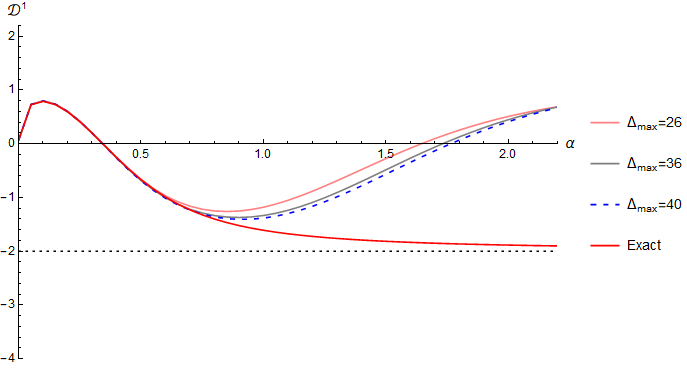}
\caption{\label{fig:your-figure5}${\cal D}^{1}_{s}(e_{1}-e_{0})$ versus $\alpha>0$ for the fixed spectator. Linear term removed using interaction matrix eigenvalues.}
\end{figure}

Using the first method and taking $s$ from the interaction matrix we obtain the  plots for each choice of spectator and different truncation 
levels presented on Fig. \ref{fig:your-figure4} and Fig. \ref{fig:your-figure5}.

\begin{figure}[H]
\centering
\includegraphics[width=0.85\textwidth]
{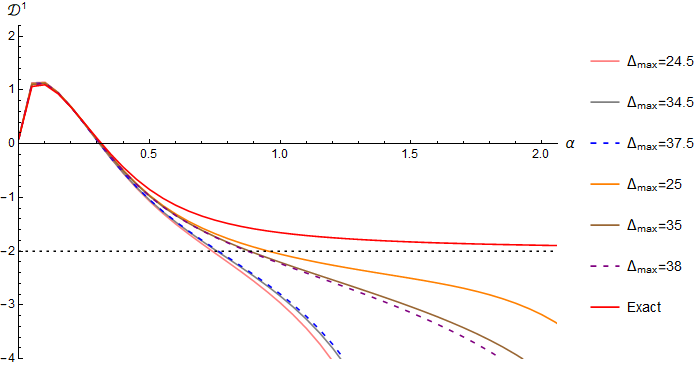}
\caption{\label{fig:your-figure6}${\cal D}^{1}_{s}(e_{1}-e_{0})$ versus $\alpha>0$ for the free spectator. Linear term removed using regression analysis 
on the interval $2<\alpha<4$.}
\end{figure}

\begin{figure}[H]
\centering
\includegraphics[width=0.85\textwidth]
{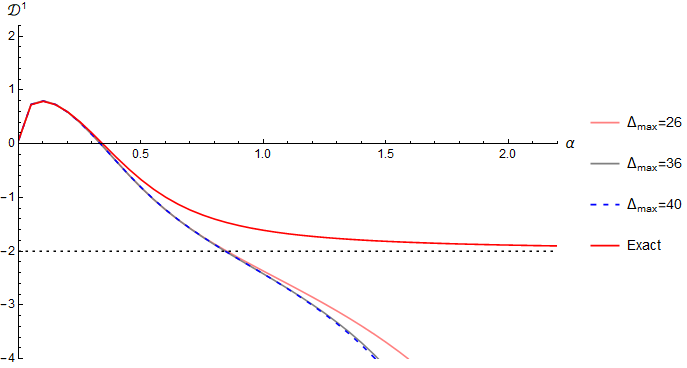}
\caption{\label{fig:your-figure7}${\cal D}^{1}_{s}(e_{1}-e_{0})$ versus $\alpha>0$ for the fixed spectator. Linear term removed using regression analysis 
on the interval $2<\alpha<4$.}
\end{figure}

For the fixed spectator and for integer $\Delta_{\rm max}$ in the free spectator case we see that the curve for ${\cal D}^{1}_{s}(e_{1}-e_{0})$ has a minimum descending towards $t=-2$ as we increase $\Delta_{\rm max}$. For the free spectator the curves for half-integer $\Delta_{\rm max}$ form a separate 
band and instead of a minimum they have an inflection point.  
Increasing  the coupling past the minimum or inflection point  we arrive to the region approximately  between $\alpha=2$ and $\alpha=4$ (not displayed in the above plots) where the curve flattens again around the value $t=1$.
The same flattening near the value $t=1$ also occurs for the fixed spectator and for the free spectator with integer $\Delta_{\rm max}$.
This  points to an approximately linear regime past the appearance of the IR fixed point. If we read off the slope numerically (via regression analysis) 
in that region and use it for the value of $s$ we obtain the  plots presented on Fig. \ref{fig:your-figure6} and \ref{fig:your-figure7}.

The region where the  curve for ${\cal D}^{1}_{s}(e_{1}-e_{0})$  flattens now looks like a region near an infection point for both spectators and all truncation levels. These regions are closer to the theoretical value $t=-2$ than with the previous choice of $s$. In section \ref{beyond_general} we discussed why such a linear regime occurs past the closest approach to the IR fixed point. 

\begin{figure}[H]
\centering
\includegraphics[width=0.85\textwidth]
{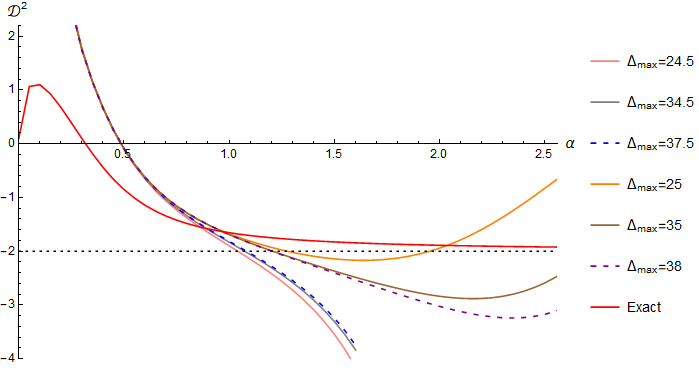}
\caption{\label{fig:your-figure8}${\cal D}^{2}(e_{1}-e_{0})$ versus $\alpha>0$ for the free spectator}
\end{figure}

For the second method that uses (\ref{d2}) we first present the results for the free spectator on Fig. \ref{fig:your-figure8}.


Again we observe a qualitatively different behaviour for integer and non-integer $\Delta_{\rm max}$. For  integer 
$\Delta_{\rm max}$ the  ${\cal D}^{2}(e_{1}-e_{0})$ curves have an approximate inflection point. These inflection points   are not 
easily discerned from the plots but are found numerically and  become more pronounced on increasing the truncation.  
Also they have a minimum for larger values of $\alpha$. It is interesting to note that such qualitative differences are not visible when looking at the $e_{i}-e_{0}$ curves. 

\begin{figure}[H]
\centering
\includegraphics[width=0.8\textwidth] 
{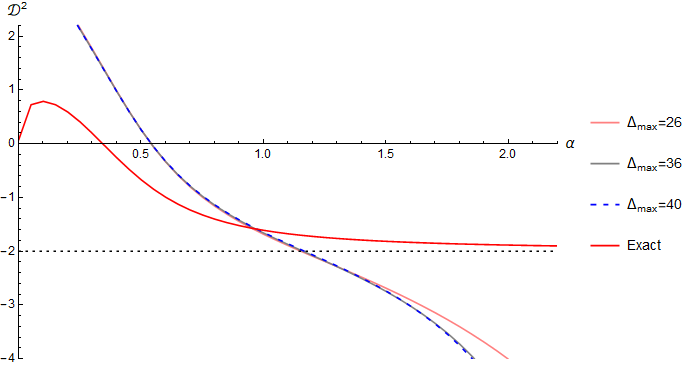}
\caption{\label{fig:fixed_spec}${\cal D}^{2}(e_{1}-e_{0})$ versus $\alpha>0$ for the fixed spectator}
\end{figure}

For the fixed spectator the plot is presented on Fig. \ref{fig:fixed_spec}.

For all three methods we observe a flattening of the ``treated gaps" ${\cal D}^{1}_{s}(e_{1}-e_{0})$ and ${\cal D}^{2}(e_{1}-e_{0})$ 
at around $\alpha \approx 1$. 
Finding the values at the minima or the inflection points we obtain best fits to the value of the leading infrared exponent. These best 
fits are tabulated below for each choice of spectator, method and a sample of truncations\footnote{The best fit for ${\cal D}^2$ for the free spectator with $\Delta_{\rm max}=25$ was calculated using the location of the minimum.  For the higher truncation levels, the location of the inflection point was used.}.
\begin{table}[h!]
\centering
\resizebox{\textwidth}{!}{
\begin{tabular}{|p{3cm}|c|c|c|c|c|c|}
\hline
\emph{} & \multicolumn{6}{c|}{$\Delta_{\rm max}$}  \\
\hline
\emph{method}  & 24.5 &34.5 &37.5 &25&35& 38 \\     
\hline
  ${\cal D}^{1}_{s}$ with $s$ from interaction matrix eigenvalues&(0.65,-1.76)&(0.64,-1.68)&(0.64,-1.68)&(0.65,-0.86)&(0.68,-0.96)&(0.68,1.00) \\
 \hline
 ${\cal D}^{1}_{s}$ with $s$ from  regression analysis&(0.713,-1.89)&(0.727,-1.89)&(0.73,-1.89)&(1.361,-2.41)&(1.133,-2.43)&(1.07,-2.36)  \\
  \hline
  ${\cal D}^{2}$ &(0.95,-1.8)&(0.965,-1.78)&(0.97,-1.75)&(1.61,-2.2)&(1.38,-2.24)&(1.44,-2.33) \\
    \hline
 \end{tabular}
}
\caption{Best fits for the leading IR exponent $t$, for the free spectator} \label{table}
\end{table}
\begin{table}[h!] 
\centering
\scalebox{0.9}{\begin{tabular}{|p{4cm}|c|c|c|}
\hline
\emph{} & \multicolumn{3}{c|}{$\Delta_{\rm max}$}  \\
\hline
\emph{method}  &26&36&40\\  
\hline
  ${\cal D}^{1}_{s}$ with $s$ from interaction matrix eigenvalues&(0.825,-1.25)&(0.868,-1.35)&(0.88,-1.4)\\
 \hline
 ${\cal D}^{1}_{s}$ with $s$ from  regression analysis&(0.988,-2.34)&(0.89,-2.17)&(0.88,-2.12)  \\
  \hline
  ${\cal D}^{2}$ &(1.2,-2.15)&(1.129,-1.94)&(1.118,-1.92) \\
    \hline
 \end{tabular}}
\caption{Best fits for the leading IR exponent $t$, for the fixed spectator} \label{table}
\end{table}

We see that   ${\cal D}^{1}_{s}$ with locally determined slope $s$ gives most accurate estimates of $t$.  
As we increase $\Delta_{\rm max}$ the estimates behave monotonically but do not always get closer 
to the theoretical values. The errors of these approximations can be attributed to truncation effects as well as to the deficiencies of 
the method  
which in particular include not  knowing the 
coefficient $A_{1}^{(1)}$ in (\ref{fp_ansatz1}) which, given the results, we may expect to be  small. 

For the free spectator we also tried  more general  $(N_{\rm max}^{1},N_{\rm max}^{\epsilon})$--truncation schemes. For 
small differences $|N_{\rm max}^{1}-N_{\rm max}^{\epsilon}|$ the results are roughly in the same ball park as for the standard truncation 
scheme. We will discuss the effects of these truncation schemes further in section \ref{beyond5}.

\subsection{Boundary flows in tricritical Ising model}\label{TIM1sec}
The tricritical Ising model (TIM) is the A-type M(5,4) minimal model with central charge $c=7/10$.  It is the continuum limit of the lattice spin model with lattice spins  taking values -1,0 and 1.  The model has  6 primary bulk fields $\psi_{i}$ with 6 associated conformal boundary conditions given by the Cardy construction.
Table \ref{TIM_dimensions} contains the notation for the primary states, their weights, common notations for the corresponding boundary conditions (inspired by 
the boundary spin interpretation) and the boundary primary fields that live on these boundary conditions.

\begin{figure}[h!]
\begin{center}
\begin{tikzpicture}
\draw (0,3.5) circle [radius=0.8] node {$(\!-\!)\!\oplus\!(\!+\!)$};
\draw (0,0) circle [radius=0.4] node {d};
\draw (0,-3.5) circle [radius=0.4] node {0};
\draw (-4,-3.5) circle [radius=0.4] node {$-$};
\draw (-2,-3.5) circle [radius=0.4] node {--\,0};
\draw (2,-3.5) circle [radius=0.4] node {0+};
\draw (4,-3.5) circle [radius=0.4] node {+};
\begin{scope}[thick, every node/.style={sloped,allow upside down}]
\draw[dashed] (0.8,3.5) to[out=0,in=90] node {\midarrow} (4,-3.1);
\draw[dashed] (-0.8,3.5) to[out=180,in=90] node {\midarrow}  (-4,-3.1);
\draw[red] (0.4,0) to[out=0,in=90] node {\midarrow} (4,-3.1);
\draw (0.4,0) to[out=-40,in=90] node {\midarrow} (2,-3.1);
\draw[red] (-0.4,0) to[out=180,in=90] node {\midarrow} (-4,-3.1);
\draw (-0.4,0) to[out=220,in=90] node {\midarrow} (-2,-3.1);

\draw[blue] (0,0.4) -- node {\midarrow} (0,2.7);
\draw[blue] (0,-0.4) -- node {\midarrow} (0,-3.1);
\draw[blue]  (-2.4,-3.5)-- node {\midarrow} (-3.6,-3.5);
\draw[blue] (-1.6,-3.5) -- node {\midarrow} (-0.4,-3.5);
\draw[blue]  (1.6,-3.5)-- node {\midarrow} (0.4,-3.5);
\draw[blue]  (2.4,-3.5)-- node {\midarrow} (3.6,-3.5);
\end{scope}
\end{tikzpicture}
\caption{\label{fig:your-figure}The space of boundary flows in the Tricritical Ising Model}
\label{space_of_flows}
\end{center}
\end{figure}
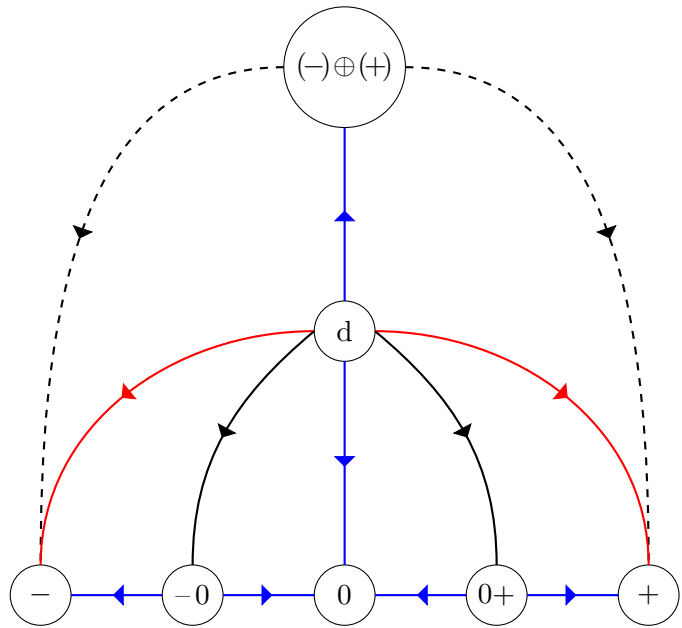

\begin{table}[H]
\centering
\begin{tabular}{|l|l|l|l|l|l|l|}
\hline
Primary Operator & $\mathbb{I}$ & $\epsilon$                       & $\epsilon^{\prime}$                      & $\epsilon^{\prime\prime}$              & $\sigma$ & $\sigma^{\prime}$                                                              \\ \hline
Virasoro Label                 & $(11)$ & $(12)$ & $(13)$ & $(31)$ & $(22)$ & $(21)$\\ \hline
Conformal Weight $\Delta$ & 0            & 1/10                             & 3/5                                   & 3/2                              & 3/80                       & 7/16                                                                  \\ \hline
Notation for Cardy b.c.            & (-)          & (-0)                             & (0+)                                    & (+)                             & (d)                       & (0)                                                                 \\ \hline
Boundary Fields           & $\mathbb{I}$ & $\mathbb{I}$,$\epsilon^{\prime}$ & $\mathbb{I}$,$\epsilon^{\prime}$ & $\mathbb{I}$& $\mathbb{I}$,$\epsilon$,$\epsilon^{\prime}$,$\epsilon^{\prime\prime}$               &$\mathbb{I},\epsilon^{\prime\prime}$ \\ \hline
\end{tabular}
\caption{Table of notation for the tricritical Ising model} \label{TIM_dimensions}
\end{table}
The space of boundary flows that start out from Cardy boundary conditions in this model is well known in the literature \cite{Gerard2},\cite{Affleck} and is depicted on Figure \ref{space_of_flows}.    
 The blue  lines on the picture represent the perturbations by $\psi_{13}$ that we will consider in this paper. The 
red lines going from $(d)$ to $(+)$ and $(-)$ boundary conditions correspond to the RG flows given by $\psi_{12}$ perturbations and 
the dashed lines stand for component flows from the $(+)\oplus(-)$ superposition of two boundary conditions. The black lines 
correspond to two flows from $(d)$ to $(-0)$ and $(0+)$  generated by a particular linear combination of $\psi_{12}$ and $\psi_{13}$ 
which so far has not been found numerically but should be there for continuity reasons. 

Some of the flows depicted on  Figure \ref{space_of_flows} are related by the action of topological defects by virtue of the Graham-Watts theorem 
\cite{GW}.  Topological defects $X_{i}$ in a Virasoro minimal model are labeled by a primary state label $i$. They act on Cardy boundary 
states $|j\rangle\!\rangle$ by means of fusion rules coefficients $N_{ij}^{k}$ so that 
\be
X_{i}|j\rangle\!\rangle = \sum_{k} N_{ij}^{k} |k\rangle\!\rangle  \, .
\ee
In particular for the tricritical Ising model the defect $X_{\epsilon''}$ is the spin-reversal defect. Its action reflects the  flows 
on  Figure \ref{space_of_flows} about the vertical line, exchanging the $+$ and $-$ labels of the boundary conditions. 
Another useful defect is $X_{\sigma'}$ which maps the horizontal pair of flows from $(0+)$ to $(+)$ and to $(0)$ into the vertical 
pair of flows from $(d)$ to $(0)$ and to $(+)\oplus(-)$. Thus to describe $\psi_{13}$-flows we can focus on the pair of flows that starts from 
$(0+)$:
\begin{equation}\label{13flows}
(0)  \longleftarrow  (0+)  \longrightarrow  (+) \, .
\end{equation}
For these flows the dimensionless Hamiltonian on a strip is given by
\begin{equation}\label{stripHam} 
h=(L_0-c/24)+\mu L  \psi_{13}(0,0)
\end{equation}
where $\mu$ is the dimensionful coupling and $\psi_{13}(0,0)$ stands for the primary field inserted on the strip at $(\tau, \sigma)=(0,0)$. 
The matrix elements of the perturbing operator can be calculated using the conformal mapping from the upper half plane to the 
strip: $w=\frac{L}{\pi}{\rm ln} z$. Here $w=\sigma + i\tau$ is the complex coordinate on the strip and $z=x+iy$ is the complex coordinate on the 
upper half plane: $y\ge 0$. Using this we rewrite (\ref{stripHam}) as
\begin{equation} \label{ham13}
h=(L_0-c/24)+\lambda  {\pi}^{-0.4} \psi_{13}(1)
\end{equation}
where $\lambda = \mu L^{0.4}$ is the dimensionless coupling \footnote{Throughout this paper, we also use $\lambda$ to represent a model independent coupling in a more general setting.  It should be clear from the context which coupling is being referred to.} and $\psi_{13}(1)$ stands for the operator inserted on the boundary of the upper half plane.  We choose a normalisation for the operators such that the identity field appearing in the OPE with itself has coefficient 1.  The matrix elements of the perturbing operator can be computed using radial quantisation.

We assume that on the bottom edge of the strip we have the $(0+)$ boundary 
condition while on the top edge we have a choice of 6 spectator boundary conditions. Choosing a spectator with label $s$ gives us 
a Hilbert space that can be decomposed into Virasoro irreducible representations as in (\ref{Hss}).
The dimensions of the truncated state space for the tricritical Ising model with (d) spectator are displayed in the Appendix \ref{App2}.
The matrix elements are calculated by sandwiching the Hamiltonian (\ref{ham13}) between states (\ref{desc_states}) and using commutation relations between 
the Virasoro generators and $\psi_{13}(1)$ to express them in terms of boundary three point functions. The latter were calculated in \cite{Runkel}. 
The flows (\ref{13flows}) were analysed in TCSA in \cite{Gerard2} where the numerics were also compared against the TBA results. 
As expected the approach to the IR fixed point was found to be  independent of the choice of spectator boundary condition. 
Our numeric computations for the energy levels, which were performed with truncated   state spaces with dimension of the order 2,500 \footnote{The dimension is spectator dependent: the results were obtained for various levels, each corresponding to a dimension of the order 2,500 for the chosen spectator.} agree with the results of  \cite{Gerard2}.



\subsection{Approach to fixed point}
We would like to focus here on reading off the leading asymptotic exponent for the energy gap 
on approach to the infrared fixed points for the two flows in (\ref{13flows}). 
For a  positive coupling $\lambda$ the flow  $(0+)\rightarrow (+)$
 is observed whilst for negative coupling the $(0+)\rightarrow (0)$ flow is realised.  
 The boundary spectrum of the $(+)$ boundary condition contains only the identity tower and the end of the flow is believed to be dominated by the stress-energy tensor. For the negative coupling the approach to the IR fixed point $(0)$ is 
 believed to be dominated by the irrelevant field $\psi_{31} \equiv \epsilon^{\prime\prime}$ field of dimension 3/2\footnote{This claim is substantiated by the fact that for large central charge the boundary $\psi_{13}$ flows become perturbative in the  negative coupling direction 
 and the dimension of the leading irrelevant operator in the IR can be shown to be that of $\psi_{31}$.}.
Using (\ref{ti}) with $\Delta ^{UV}=3/5$ for the $\psi_{13} \equiv \epsilon^{\prime}$ field and $\Delta ^{IR}_{(+)}=2$ and $\Delta ^{IR}_{(0)}=3/2$ for the stress energy tensor and $\psi_{31}$ field we find
\be
 t_T ^{\text{TIM}}=\frac{5}{2}\, , \qquad 
 t_{\psi_{31}} ^{\text{TIM}}=\frac{5}{4}\, .
\ee
To find numerical approximations to these theoretical values we use the same methodology \footnote{The analysis is applied to the second excited state $E_2$ as this state merges with the third excited state enabling us to better predict when the fixed point is reached. The linear regression  for the operator ${\cal D}^{1}_{s}$ was performed in the region $1< \lambda < 2$.}  as in the Ising model section with numerical best fits for the $(d)$ spectator displayed in Table \ref{table_TIM_exp} below.
\begin{table}[h!] 
\centering
\resizebox{\textwidth}{!}{
\begin{tabular}{|p{2cm}|c|c|c|c|c|c|}
\hline
\emph{} & \multicolumn{3}{c|}{\emph{Positive Coupling}}  & \multicolumn{3}{c|}{\emph{Negative Coupling}} \\
\hline
\emph{method}\enspace \enspace \enspace \textbackslash  \hspace{0.9cm} $\Delta_{\rm max}$ &16&18&20&16&18&20\\
\hline
  ${\cal D}^{1}_{s}$ with $s$ from interaction matrix eigenvalues &(0.641,-1.05)&(0.644,-1.11)&(0.65,-1.17)&(0.208,-1.37)&(0.211,-1.43)&(0.214,-1.55)\\
 \hline
 ${\cal D}^{1}_{s}$ with $s$ from  regression analysis&(0.704,-2.54)&(0.697,-2.54)&(0.69,-2.54)&(0.54,-1.63)&(0.53,-1.57)&(0.52,-1.55)\\
  \hline
  ${\cal D}^{2}$ &(0.975,-2.62)&(0.946,-2.56)&(0.925,-2.5)&(0.702,-1.43)&(0.7,-1.37)&(0.69,-1.31)\\
    \hline
 \end{tabular}
 }
\caption{Best fits for the leading IR exponent $t$ for positive and negative coupling} \label{table_TIM_exp}
\end{table}

\begin{figure}[H]
  \begin{subfigure}[b]{0.47\textwidth}
    \includegraphics[width=\textwidth]{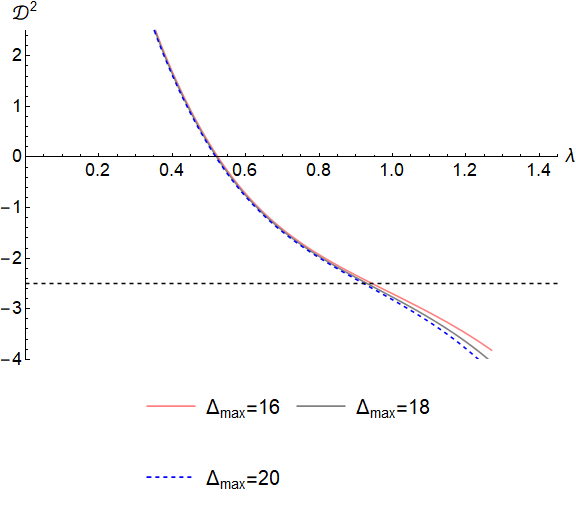}
    \caption{${\cal D}^{2}(e_{2}-e_{0})$ versus $\lambda>0$ in the TIM with (d) spectator boundary condition}
  \end{subfigure}
  \qquad	
  \begin{subfigure}[b]{0.47\textwidth}
    \includegraphics[width=\textwidth]{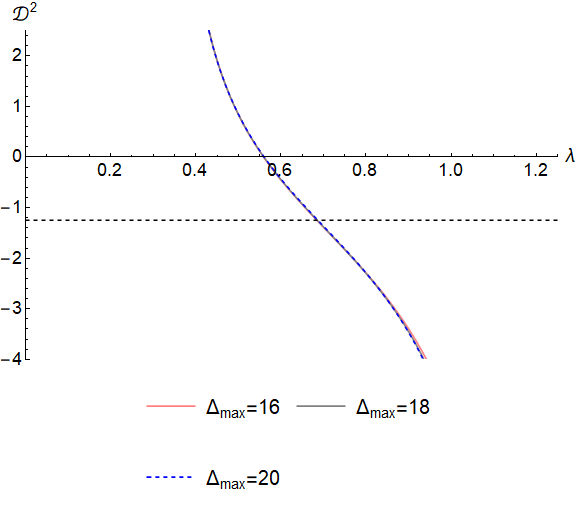}
    \caption{${\cal D}^{2}(e_{2}-e_{0})$ versus $\lambda<0$ in the TIM with (d) spectator boundary condition}
  \end{subfigure}
  \caption{} \label{D2_TIM_plots}
\end{figure}

On Figure \ref{D2_TIM_plots} we present the plots for the third method based on (\ref{d2}). For brevity 
we present only the results for  the spectator boundary condition given by the Cardy state  $|\sigma\rangle\!\rangle$.  We see similar behaviour near the fixed point as in the Ising model where we have an inflection point in similar plots. 
 Increasing the truncation weight  improves the approximation of the leading power bringing it closer to the theoretical value (5/2 for positive coupling and 5/4 for negative coupling).
We note that similar behaviour is observed for the other spectators.


\section{Flows beyond the IR fixed point} \label{FlowsBeyond_section}
\setcounter{equation}{0}
\subsection{Examples}\label{beyond_examples}
In the examples discussed in the previous section  as we increase the coupling past the IR fixed point 
we observe a rearrangement of multiplicities characteristic of a different fixed point. We call these ``flows beyond the 
IR fixed point'', or just ``flows beyond''. They were observed and discussed in \cite{Toth},\cite{Feverati_etal},  \cite{Gerard2}.

\begin{figure}[H]
  \begin{subfigure}[b]{0.47\textwidth}
    \includegraphics[width=\textwidth]{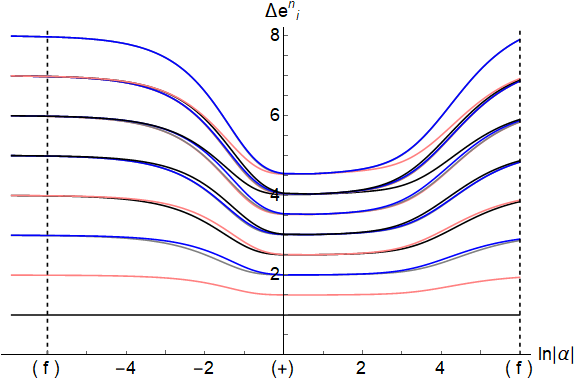}
    \caption{$ \Delta e_{i}^{\rm n}$  versus ${\rm ln}|\alpha|$ for $\alpha>0$, $\Delta_{\rm max}=34$ for the fixed spectator}
  \end{subfigure}
  \qquad	
  \begin{subfigure}[b]{0.47\textwidth}
    \includegraphics[width=\textwidth]{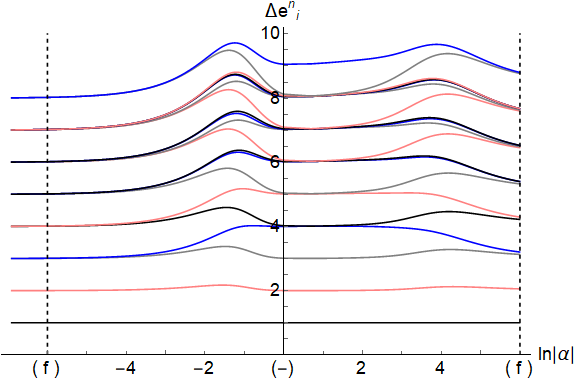}
    \caption{$\Delta e_{i}^{\rm n}$  versus ${\rm ln}|\alpha|$ for $\alpha<0$, $\Delta_{\rm max}=34$ for the fixed spectator}
  \end{subfigure}
  \caption{}
  \label{RamReverseFlows}
\end{figure}

The simplest case of a flow beyond appears to be the boundary magnetic field flow with a fixed spin spectator. 
If we are to plot the energy gaps $e_{i}-e_{0}$ versus the coupling  we will see that past the fixed point 
they settle asymptotically into a linear regime\footnote{This is the second linear regime past the $\lambda=\lambda^{**}$ threshold discussed in section \ref{beyond_general}.}. We can interpret this linearity as due to energy rescalings stemming from truncation effects. Such rescalings have been discussed in the perturbative regime in \cite{Gerard2}. Looking 
instead at the normalised energy gaps: 
\be
\Delta e_{i}^{\rm n} = \frac{e_{i}-e_{0}}{e_{1}-e_{0}} \, . 
\ee
we remove the energy rescalings and can observe the 
asymptotic multiplicities rearranging themselves (here for simplicity we assume that the vacuum does not become asymptotically degenerate). The normalised gaps for the flow at hand  are presented on Figure \ref{RamReverseFlows} using the logarithmic scale ${\rm ln}|\alpha|$ for ease of visualisation.
These plots show that for very large absolute values of the coupling the multiplicities rearrange back to those of the original 
UV fixed point so that the flow beyond looks like the reverse of the physical  flow from the  UV free boundary condition to the 
IR fixed boundary condition.  In the plots we have marked with dashed lines approximate regions representing  the 
vicinities of fixed points (the IR fixed point occurs approximately along the $\alpha =0$ axis). Of course the flow beyond is not physical in that it violates the $g$-theorem\cite{gThm1, gThm2}.  It does not happen in the continuum theory and is entirely an artefact of the truncation. However, we find this phenomenon interesting in that it 
 does arise in a variety of situations when using TCSA, especially for boundary perturbations, and understanding the mechanism behind it  must  tell us something about how TCSA works at large couplings.  

In general, in TCSA we are dealing with a Hamiltonian of the form 
\be
h=L_{0}^{\rm UV} + \lambda V^{\rm UV}
\ee
where $L_{0}^{\rm UV}$ and $V^{\rm UV}$ are finite matrices.
 Asymptotically, possibly after a number of different intermediate regimes, the interaction 
matrix is bound to dominate over the $L_{0}^{\rm UV}$ matrix and the energy gaps will be given by the difference of the eigenvalues 
$v_{i}$ of the interaction matrix $V^{\rm UV}$ times the coupling. More precisely,  assuming the eigenvalues $v_i$ are non-degenerate and 
$v_{\rm min}$ is the smallest eigenvalue and  $v_{\rm max}$ is the largest eigenvalue  we have 
\be
e_{i}-e_{0} \sim \lambda (v_{i}-v_{\rm min})  \enspace \mbox{when}\enspace \lambda \to \infty\, ,  
\ee
\be
e_{i}-e_{0} \sim \lambda (v_{i}-v_{\rm max})  \enspace \mbox{when} \enspace \lambda  \to -\infty \, .
\ee
For the normalised gaps we then get an asymptotic approach to constant values 
\be \label{delta_e_as}
\Delta e_{i}^{\rm n} \sim \frac{v_{i} - v_{\rm min}}{v_{1}-v_{\rm min}} \enspace \mbox{when}\enspace \lambda \to \infty 
\ee
and similarly for $\lambda \to -\infty$. 
For degenerate eigenvalues of $V^{\rm UV}$ the situation is a bit more interesting. If $e_{i}$ and $e_{j}$ are two 
eigenvalues of $h$ that asymptotically tend to the same degenerate eigenvalue, $v_k$, then their difference 
will be given asymptotically by the leading correction, which is equal to an eigenvalue of the operator $L_{0}^{\rm UV}$ 
reduced to the eigenspace for $\vec{v}_{k}$. This will affect the asymptotic behaviour (\ref{delta_e_as}) 
when the smallest (or largest) eigenvalue of $V^{\rm UV}$ is degenerate. 

Coming back to the boundary magnetic field flow, we checked that the smallest and largest eigenvalues of $V^{\rm UV}$, 
which differ only by the sign, are non-degenerate for the fixed spin spectator. The asymptotic behaviour (\ref{delta_e_as})  
holds numerically to good accuracy. The flow beyond is simply described (asymptotically) as the dominance of the 
interaction matrix. However the multiplicities for low lying eigenvalues of $V^{\rm UV}$ turn out to be the same as those of 
$L_{0}^{\rm UV}$. For a fixed number of low lying states it is essential 
to take large enough $\Delta_{\rm max}$ to observe this as for small $\Delta_{\rm max}$ the multiplicities still deviate. 
We have no simple explanation of this coincidence.

Although, to avoid clutter, in Fig \ref{RamReverseFlows} we present only the normalised gaps obtained for a particular value of $\Delta_{\rm max}$ 
we have made a comparison for a number of  different values of $\Delta_{\rm max}$ and observed that the value of $\alpha$ 
at which we enter the asymptotic regime with the UV BCFT multiplicities  increases as we increase  $\Delta_{\rm max}$. 
Physically this supports the intuition  that flows beyond constitute a truncation effect and are absent in the continuum theory. 
 Qualitatively this also ties in with the domination of $\alpha V^{\rm UV}$ matrix that would set in for larger values of $\alpha$ 
because the largest eigenvalue of  $L_{0}^{\rm UV}$ grows linearly with $\Delta_{\rm max}$  while that of $V^{\rm UV}$ only grows as a square 
root of $\Delta_{\rm max}$. 

\begin{figure}[H]
\centering
\includegraphics[scale=0.5]{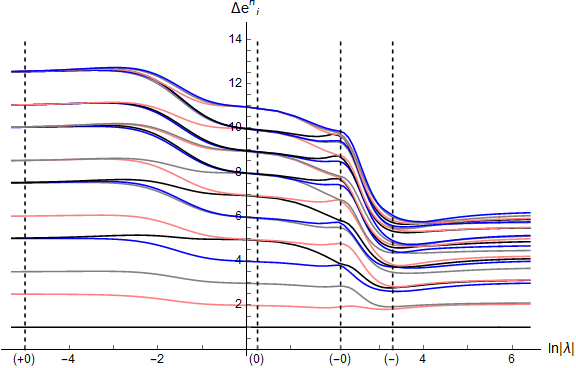}
\caption{\label{fig:your-figure} $\Delta e_{i}$ versus ${\rm ln}|\lambda|$  in the TIM with $(d)$-spectator and $\lambda <0$ with $\Delta_{\rm max}=20$} \label{TIM_beyond_plots}
\end{figure}

 The phenomenon of flows beyond is even more spectacular for the boundary $\psi_{1,3}$ flows in the TIM. 
As explained in section \ref{TIM1sec} due to the Graham-Watts theorem we can focus on the $\psi_{13}$-flows originating from the $(0+)$ boundary condition. While for positive values of the coupling we find a physical RG flow to the $(+)$ boundary condition and no flow beyond, 
for negative coupling we find a cascade of flows beyond that can be summarised by the following diagram
\begin{equation}
(0+) \rightarrow (0) \rightarrow (-0) \rightarrow (-) 
\end{equation}
as discerned from the changes in  multiplicities of the normalised energy gaps plotted on Figure \ref{TIM_beyond_plots}.
In these  plots we chose  the disordered boundary condition $(d)$ as a spectator but, remarkably, the cascade of multiplicity changes is the same 
for any choice of 6 spectator boundary conditions. These flows beyond were first investigated in \cite{Gerard2}. 
While the far end of the cascade can be still described by the dominance of the interaction matrix $V^{\rm UV}$ within a degree of numerical accuracy, particularly in the low lying spectrum, the intermediate 
linear regimes with rearranged multiplicities cannot be described so simply. The independence from the spectator boundary condition suggests 
some local description in the space of effective Hamiltonians of the type discussed in section \ref{beyond_general}. In that description the 
flows beyond are described by perturbations of IR fixed points by irrelevant operators with flipped sign. Crucially, to determine 
where the bounce flow is headed next, depends on the truncation scheme at the IR fixed point. In the next subsection we describe two 
exactly solvable models of truncated perturbations that offer further insights supporting the picture outlined in section  \ref{beyond_general}.

\subsection{Two exactly solvable models} \label{exact_beyond}
In this section, we consider two analytically solvable models with the same truncation scheme.  The first example 
shows that by using an alternative scheme to TCSA, there may be no flow beyond the IR fixed point.  The second is an example of a theory with
a bounce flow -- flowing first to a new fixed point, then returning back to the original one. 

In the first example we use a mode truncation scheme for the boundary magnetic field flow in the Ising model where we leave the spin to fluctuate freely on the spectator boundary. 
The mode truncation scheme for this flow was first considered in \cite{Toth} where it was noted that there is no flow beyond. 

For the free spectator the Hamiltonian in the mode truncation scheme, written in the notation of section \ref{Ising1_sec} is 
\begin{equation}\label{NHS_mod}
H^{\rm NS}_{\rm mod} = \frac{\pi}{L}\Bigl[ \sum_{k=0}^{n_c} (k+1/2) a_{k+1/2}^{\dagger}a_{k+1/2} - \frac{1}{48}
+ i\alpha \sum_{k=0}^{n_c} (a_{k+1/2}^{\dagger} + a_{k+1/2})a \Bigr]
\end{equation}
restricted to the subspace of the physical space spanned by the states  of the form (\ref{NSbasis}) built with 
operators $a_{i+1/2}^{\dagger}$, $i=0, \dots, n_c$. The integer $n_c$ is the mode truncation parameter. The dimension of 
the truncated state space is $2^{n_c+1}$. The Hamiltonian (\ref{NHS_mod}) can be diagonalised by means of a Bogolyubov transformation.
To find the latter we write an ansatz for the diagonalising modes:
\be \label{B1}
b_{\omega} =  \sum_{k=0}^{n_c} (A_{\omega, k}a^{\dagger}_{k+1/2} + B_{\omega, k}a_{k+1/2}) + \frac{a}{f(\omega)}\, ,
\ee
and require that it satisfies 
\be \label{com_eq}
[ H_{\rm mod}^{\rm NS}, b_{\omega}] = \omega b_{\omega} \, .
\ee
This results in a  system of linear equations on the coefficients $A_{\omega, k}$, $B_{\omega, k}$, $f(\omega)$ that has  a non-trivial solution 
provided $\omega$ satisfies the following  equation
\be \label{spec_eq_mod}
\sum_{k=0}^{n_c}\frac{2\omega^2}{(k+1/2)^2 - \omega^2} = -\frac{\omega^2}{2\alpha^2} \, .
\ee
This  equation has $n_c+1$ bounded modes which include zero and  $n_c$ solutions 
$\omega =\pm  \tilde \omega_{n}$, $\tilde \omega_{n}>0$ that remain bounded as $\alpha \to \infty$. It also has one heavy mode solution 
$\omega=\Omega_{\alpha}$ where 
\be
\Omega_{\alpha} \sim 2|\alpha| \sqrt{n_c + 1} \enspace \mbox{ as } \alpha \to \pm \infty \, .
\ee 
The state space then splits neatly into two subspaces: the one with light states that is built using $b^{\dagger}_{\tilde \omega_{n}}$ and the one 
with  heavy states built using $b^{\dagger}_{\Omega_{\alpha}}$ and other oscillators. In the heavy states subspace the energy gaps
 diverge linearly as $\alpha \to \pm \infty$. Focussing on the light states one has 
 \be
\lim_{\alpha \to \pm \infty} \tilde \omega_{n} = n + {\cal O}\left( \frac{1}{n_c}\right)\, ,   \quad n=1,\dots, n_c 
 \ee
so that up to  errors suppressed by the truncation parameter we obtain the spectrum of the fixed spin boundary condition describing 
the IR fixed point. The vacuum energy also diverges linearly in $\alpha$ due to the $\Omega_{\alpha}$ mode contribution. In the 
energy gaps $e_{i}-e_{0}$ this divergence cancels for the light states.  
We can interpret this asymptotic behaviour in terms of the dominance of the interaction matrix $V$. The latter has only two eigenvalues: 
$\sqrt{n_c+1}$ and $-\sqrt{n_c+1}$, with equal dimensions. The light states including the vacuum are, asymptotically, perturbations of 
the states in the  $-\sqrt{n_c+1}$-subspace for $\alpha \to \infty$ and of the $\sqrt{n_c+1}$-subspace for $\alpha \to -\infty$. 
While the states in the heavy states subspace  all decouple, the light states give the approximate spectrum of the IR fixed point due 
to the first perturbative correction.  In a bit more detail we write 
\be
L_{0}^{\rm UV} + \alpha V^{\rm UV} = \alpha ( V^{\rm UV} + \frac{1}{\alpha}L_{0}^{\rm UV}) 
\ee
and treat $\frac{1}{\alpha}L_{0}^{\rm UV}$ as a perturbation. In the energy gaps the leading terms from the appropriate eigenvectors 
of $V^{\rm UV} $ cancel each other while the first order correction from the perturbation gives rise to constant terms in the 
$\alpha \to \pm \infty$ limit. The numerical values according to the secular equation are the eigenvalues of  $L_{0}^{\rm UV}$ 
restricted to the appropriate eigensubspace of $V^{\rm UV} $. It is remarkable that up to corrections suppressed by $1/n_c$ 
this gives the correct spectrum of $L_{0}^{\rm IR}$. We are not aware at present of any simple explanation of this fact. 

For the fixed spectator the story is very much similar. The mode truncated Hamiltonian reads 
\begin{equation}\label{RHam2}
H^{\rm R}_{\rm mod} = \frac{\pi}{L}\Bigl( \sum_{n=1}^{n_c} nb_n^{\dagger}b_n + \frac{1}{24}  + i\alpha [\sum_{n=1}^{n_c} (b_n^{\dagger} + b_n) + b_0]a \Bigr)
\end{equation}
and the spectral equation (\ref{spec_eq_mod}) is replaced by 
\be
\sum_{k=1}^{n_c}\frac{2\omega^2}{k^2-\omega^2} - 1 =  -\frac{\omega^2}{2\alpha^2} \, .
\ee
As we increase the coupling the elementary low lying spectrum changes from integers to approximately half-integers, with corrections suppressed 
by $n_c$. No flow beyond takes place. 

Our second model is the boundary perturbation of the Ising model by the stress-energy tensor restricted to the boundary: 
 $T=-\frac{L}{2\pi}\! :\! \!\psi\partial_{\tau} \psi\!\!:\!\!(0,0)  $  with a coupling $g$. Choosing the free spin spectator and  using the mode truncation scheme 
  this model can be described by  the following dimensionless Hamiltonian 
\bea \label{halphag}
h_{g}^{n_c}   = &&  - \frac{1}{48}+ \sum_{k=0}^{n_c}(k+\frac{1}{2})a^{\dagger}_{k+1/2}a_{k+1/2}   \nonumber \\
&& - \frac{g}{2}:\!\left(\sum_{k=0}^{n_c}\sum_{l=0}^{n_c}( a^{\dagger}_{k+1/2} +a_{k+1/2})l(a^{\dagger}_{l+1/2} - a_{l+1/2})\right)\!: \, .
\eea
The spectral equation is
\be \label{g_spec}
\sum_{k=0}^{n_c}\frac{2\omega^2}{(k+1/2)^2 - \omega^2} = -\frac{(1+g(n_c+1))^2}{ g(1+\frac{g}{2}(n_c+1)) }  \, .
\ee
Being an irrelevant coupling, $g$ goes to zero along the RG flow. If however we investigate the spectrum as we change $g$ from zero 
to plus or minus infinity, thus going backwards along the flow, we observe a curious pattern. 
For positive values of $g$, solutions to (\ref{g_spec}) are comprised  of light frequencies with values $\omega_{k} \approx k+1/2$ 
and a single heavy frequency $\Omega_{g}$ that for large $g$ is given by
\be
\Omega_{g} \approx |g| \left( \frac{n_c+1}{\sum_{k=0}^{n_{c}}(k+1/2)^2}\right)^{1/2} \, .
\ee
The large $g$ low lying spectrum is described by the equation 
\be \label{g_spec1}
\sum_{k=0}^{n_c}\frac{2\omega^2}{(k+1/2)^2 - \omega^2} = -2(n_c+1)
\ee 
that up to corrections of order $1/n_c$ is given by half integers. This means that the excitation spectrum is changed by terms suppressed by $n_c$.
 The low lying spectrum  practically does not flow. 

For negative values of $g$ we pass through a zero of the function on the right hand side of (\ref{g_spec}) at $g=-1/(n_c+1)$ and then a pole  
at $g=-2/(n_c+1)$. Increasing $g$ in absolute value we thus first arrive at a point with low lying spectrum described in terms of integer modes 
and then back to the same spectrum of the starting point $g=0$. This actually fits nicely  with the absence of flow beyond in the boundary magnetic 
field perturbation regulated the same way (with a fixed spectator). We know that in that flow we approach the infrared fixed point along the $T$-perturbation with a negative  
coupling. Following the empirical rule  that near the infrared fixed point the incoming flows get reflected, we depart the (continuum) infrared fixed point along the 
$T$-perturbation with a positive coupling. But, as we have seen before, this inflicts only minor corrections to the IR spectrum which are suppressed by 
$n_c$.  Note that to observe the asymptotic multiplicities of the flows beyond we do not need to resort to the normalised gaps $\Delta e_{i}^{\rm n}$ as no energy rescalings seem to be generated by this truncation scheme. For the first model considered in this section this was noted in \cite{Toth}.

For the fixed spectator the mode truncated $T$-perturbation is described by 
\be\label{g_Ham2}
\tilde h^{n_c}_{g} = \frac{1}{24} + \sum_{k=1}^{n_{c}}k\, b^{\dagger}_{k}b_{k} 
- \frac{g}{2}:\!\left(\left[ \sum_{k=1}^{n_c} (b^{\dagger}_{k} +b_{k}) + b_{0}\right] \sum_{l=1}^{n_c} l(b^{\dagger}_{l} - b_{l})\right)\!: \, .
\ee
It can be diagonalised by a Bogolyubov transformation that gives a spectral equation 
\be\label{g_spec2}
\sum_{k=0}^{n_c}\frac{2\omega^2}{n^2 - \omega^2} -1=-\frac{(1+\frac{g}{2}(1+2n_c))^2}{g(1+g(1+2n_c)/4)} \, .
\ee
This equation has the same qualitative features as (\ref{g_spec}) with a bounce flow in the negative $g$ direction and a ``no flow" 
in the positive $g$ direction.

It is worth mentioning that in the continuum theory perturbations by $T$ have a very different character depending on the sign of the coupling. The operator $T$ restricted to a conformal boundary is proportional to the displacement  operator. 
Perturbing  with a negative coupling is equivalent to displacing the boundary outwards, away from the bulk while perturbing with a positive coupling is equivalent to displacing the boundary inwards.
Consider  the partition function on a  cylinder of length $L$ and circumference $R$ with the perturbed boundary condition imposed on both ends, we have two possible quantisations which in string theory are usually referred to as the closed and the open string channels.  In the open string channel we represent the 
partition function in terms of the boundary (strip) spectrum while in the closed string channel we use a representation of the boundaries in terms of boundary states $\ket{\alpha}\!\rangle$:
\begin{equation}
\mathcal{Z}=\langle\!\bra{\alpha}e^{-LH_{\text{cyl}}^{\sigma}}\ket{\alpha}\!\rangle
\end{equation}
where   
\begin{equation}
H_{\text{cyl}}^{\sigma}=\frac{2\pi}{R}(L^{\sigma}_0 - c/24)
\end{equation}
is the Hamiltonian on the cylinder generating translations in the $\sigma$ direction. 
The boundary state $\ket{\alpha}\!\rangle$, written as the T-perturbation of, say, the fixed boundary condition, reads
\begin{equation}
\ket{\alpha}\!\rangle=e^{\frac{2\pi \tilde g}{R}L_0^{\sigma}}|+\rangle\!\rangle
\end{equation}
where $\tilde g$ is the dimensionful coupling. 
For a positive $T$-perturbation, when the boundaries move towards each other 
we obtain a singularity in the partition function at the value of dimensionless coupling
\begin{equation}
g_{\rm crit}=\frac{\tilde g_{\rm crit}}{L}=\frac{1}{2} 
\end{equation}
when the unperturbed boundaries sit on top of each other. This clearly signals a pathology of the boundary spectrum 
at this value of the coupling.

For negative values of the coupling the two ends of the cylinder are shifted away from each other and there is no singularity. This behaviour is qualitatively similar to the $T\bar T$ perturbation of the bulk theories discussed in \cite{Zam_TT}, \cite{Tateo_etal}. 

We should emphasise that in considering the $T$-perturbation in this section we solve the truncated version of the theory, not adding 
any counter terms which would be needed in a continuum limit. Such regulated theories should then be very sensitive to the 
method of regularisation. In the next section we take up the $T$-perturbation in TCSA. It shows a qualitatively different behaviour from the mode truncated case
for positive couplings. 

\subsection{Perturbations by irrelevant operators in TCSA} \label{beyond3}

As  discussed in section \ref{beyond_general} we assume  that the TCSA Hamiltonian in the vicinity of an IR fixed point can be described by a continuum theory with a tower of irrelevant operators switched on. Under certain assumptions the flow towards a fixed point gets reflected in the irrelevant operators space of couplings. In the models we studied numerically the flows beyond eventually terminate with a ``no flow" 
(that is, small changes to the spectrum are suppressed as we increase $\Delta_{\rm max}$). 
Presumably this can also be modelled as some particular perturbation of the fixed point by irrelevant operators. The exactly solvable case of the $T$-perturbation in the mode truncation scheme shows this. We would like to model this situation using TCSA numerics, that is, find particular perturbations 
by irrelevant operators that in TCSA demonstrate approximately no flow.  

\subsubsection{$T$-perturbations}
\begin{figure}[H]
  \begin{subfigure}[b]{0.47\textwidth}
    \includegraphics[width=\textwidth]{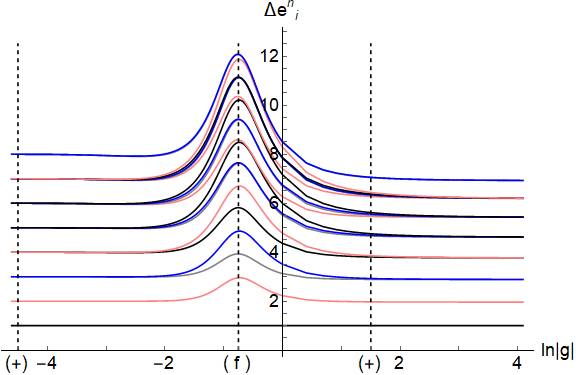}
    \caption{T perturbation of the Ramond Fixed point with positive coupling with $\Delta_{\rm max}=26$}
   \end{subfigure}
  \qquad	
  \begin{subfigure}[b]{0.47\textwidth}
    \includegraphics[width=\textwidth]{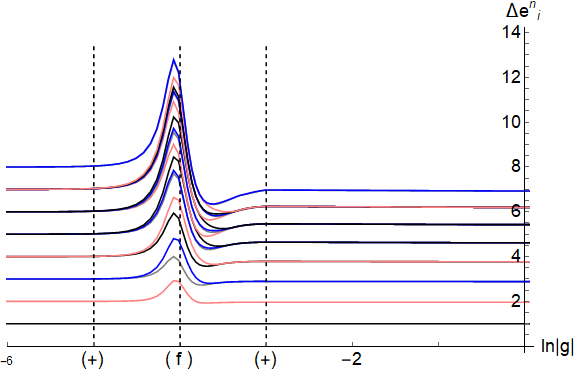}
    \caption{T perturbation of the Ramond Fixed point with negative coupling with $\Delta_{\rm max}=26$}
   \end{subfigure}
  \caption{}
  \label{fig14}
\end{figure}

We first consider in the Ising model the case when the IR fixed point is described in terms of Ramond fermions. 
This corresponds to  having a fixed  boundary condition on the flowing end of the strip and a free one on the spectator end. 
The leading irrelevant operator along which the boundary magnetic field flows terminate is the stress-energy tensor, $T$. The corresponding perturbed  Hamiltonian is
\begin{equation}\label{T pert Ramond}
h^{IR} = \frac{1}{24} + \sum_{k=1}^{\infty}k\, b^{\dagger}_{k}b_{k} 
- \frac{g}{2}:\!\left(\left[ \sum_{k=1}^{\infty} (b^{\dagger}_{k} +b_{k}) + b_{0}\right] \sum_{l=1}^{\infty} l(b^{\dagger}_{l} - b_{l})\right)\!: 
\end{equation}
with a coupling constant $g$ which we allow to vary freely. Taking a negative  $g$ and increasing it in magnitude corresponds to flowing   backwards along the boundary magnetic field flow. For large enough $|g|$  the higher dimensional irrelevant operators will become important. We will neglect these in the initial analysis. 
Plots of the normalised energy gaps for both signs of the coupling $g$ are presented below on Fig. \ref{fig14} against ${\rm ln}|g|$.

We see that going in the negative $g$ direction the spectrum describes a loop in the theory space when we first `reverse' the flow from the UV fixed point given by the free boundary condition and then flow back into the $g=0$ fixed point. Interestingly the same loop is observed when flowing into the unphysical positive $g$ direction. 
The only difference between the two directions is that for positive $g$ the vicinity of the UV fixed point is reached later than for the negative coupling. 
This is in contrast with the exact solution for the mode truncated $T$-perturbation discussed in section \ref{exact_beyond} 
where there is a ``no flow" for  positive $g$ perturbation.



We next  consider the $T$-perturbation of the IR fixed point that corresponds to having fixed boundary conditions on both ends.  The perturbed Hamiltonian is
\begin{equation}\label{T pert NS}
\begin{split}
h^{IR} =& -\frac{1}{48} + \sum_{k=0}^{\infty}(k+1/2)\, b^{\dagger}_{k+1/2}b_{k+1/2} 
\\& - \frac{g}{2}:\left( \sum_{k=0}^{\infty} (b^{\dagger}_{k+1/2} +b_{k+1/2}) \sum_{l=0}^{\infty} l(b^{\dagger}_{l+1/2} - b_{l+1/2})\right): .
\end{split}
\end{equation}

\begin{figure}[H]
  \begin{subfigure}[b]{0.47\textwidth}
    \includegraphics[width=\textwidth]{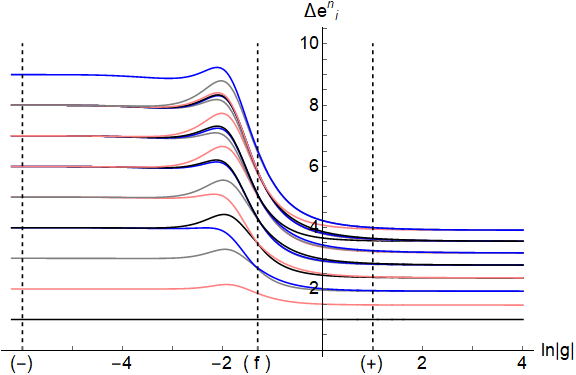}
    \caption{Perturbation of NS fermions ($\epsilon$ representation) by T with positive coupling}
    \label{fig:1}
  \end{subfigure}
  \qquad	
  \begin{subfigure}[b]{0.47\textwidth}
    \includegraphics[width=\textwidth]{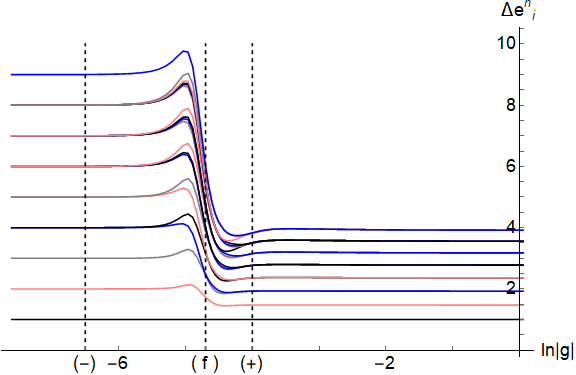}
    \caption{Perturbation of NS fermions ($\epsilon$ representation) by $T$ with negative coupling}
    \label{fig:2}
  \end{subfigure}
  \caption{}
  \label{fig15}
\end{figure}

This is described by a single primary tower in the NS fermions Fock space. 
We choose the boundary conditions below to have the fixed up boundary condition on the top end of the strip and the fixed down one on the bottom end with representation built on the Virasoro $\epsilon$-tower.  The numerical values  for normalised energy gaps are plotted on Fig. \ref{fig15}
against ${\rm ln}|g|$ for $\Delta_{\rm max}=27.5$ corresponding to a basis of 957 states. 
 
The flow generated by this perturbation is described by the following diagram
\be  \label{reverse2}
(-) \rightarrow (f) \rightarrow (+)\, .
\ee 
The spectrum   interpolates between the two NS primary towers (the identity and $\epsilon$) with an intermediary fixed point describing the free boundary condition with a fixed spectator. The fact that we flow beyond the $(f)$ 
boundary condition towards the reversed spin fixed boundary condition suggests that the relevant coupling 
of $\psi_{13}$ near $(f)$ gets reflected, which seems to be the pattern with the flows beyond. The spin reversal on the flowing end was not visible with the free spectator due to the spin reversal symmetry.  The reversal of spin in (\ref{reverse2}) 
 also means that the $T$-perturbation of the IR fixed point does not give the same loop as the flow beyond starting from the UV fixed point (which we observed for the fixed spectator in section \ref{beyond_examples}).

We also note that, as with the free spectator,  the behaviour is qualitatively different from that of the mode truncation scheme with both directions of the flow containing a further flow beyond the fixed point.
The flow is realised much earlier for negative coupling as was also seen in the previous example.

\begin{figure}[H]
  \begin{subfigure}[b]{0.47\textwidth}
    \includegraphics[width=\textwidth]{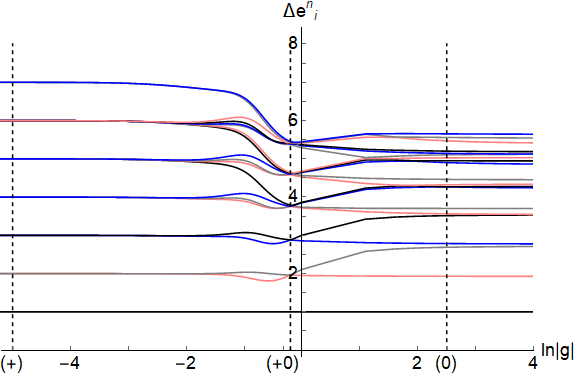}
    \caption{$T$-perturbation of the (+) boundary condition with the identity spectator in TIM,  positive coupling $\Delta_{\rm max}=25$}
    \label{fig:1}
  \end{subfigure}
  \qquad	
  \begin{subfigure}[b]{0.47\textwidth}
    \includegraphics[width=\textwidth]{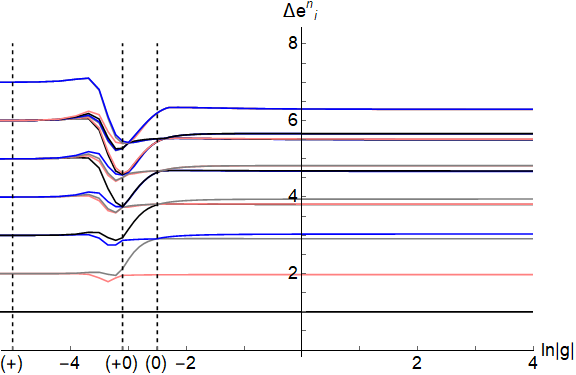}
    \caption{$T$-perturbation of the (+) boundary condition with the identity spectator in TIM,  negative coupling $\Delta_{\rm max}=25$}
    \label{fig:2}
  \end{subfigure}
  \caption{}\label{T_plots_TIM}
\end{figure}

We have also looked into the $T$-perturbation of the $(+)$ boundary condition in the tricritical Ising model. This is 
a stable infrared fixed point for the $\psi_{13}$ flow originating from the $(0+)$ boundary condition under positive coupling discussed in section \ref{beyond_examples}. This is also the end point of the cascade of $\psi_{13}$ flows from (-0) which are equivalent, by spin reversal, to flows from (0+) with negative coupling ending at (-), also discussed in \ref{beyond_examples}.
\\
\\
The Hamiltonian describing this perturbation is
\be \label{T pert TIM}
h^{\rm IR}= L_{0}^{\rm IR}  +g T(0).
\ee
On Figure \ref{T_plots_TIM} we present  plots of the normalised energy gaps 
against ${\rm ln}|g|$.

We observe that for both positive and negative coupling we move backwards along the cascade of flows beyond discussed in 
section \ref{beyond_examples}. Thus, the $T$-perturbation alone does  quite describe the end of the cascade, generating 
instead, as in the above examples, a flow back to the nearest UV fixed point first, followed by further flows beyond before stopping. In the next section we discuss how higher dimension irrelevant 
operators added to $T$ can qualitatively change the $T$-perturbation spectrum for large positive coupling in the Ising model. 

\subsubsection{Higher order irrelevant perturbations in the Ising model}

We would like  to modify the irrelevant perturbation of the IR fixed point by adding higher 
dimension operators to $T$. In doing so we try to find a situation of no flow in the reflected direction (positive coupling of $T$). The higher dimension operators we add are supposed to model, in the IR theory, the truncation effects of the UV theory as well as the subleading irrelevant operators present in the continuum theory on approach to the IR fixed point. The first operator of dimension higher than $T$ is $:\!\psi \partial_{\tau}^3 \psi\!:$. We have experimented adding it to the $T$-perturbation with different relative signs and found  significant changes in the asymptotic spectrum sensitive to the relative sign. However we did not see a no flow situation. Adding other fermionic 
bilinears does not change the situation drastically. It should be noted that all fermion bilinears are closed under OPE and thus keep the perturbed continuum theory within a certain subspace under the RG action. The smallest dimension operator that is not bilinear in the fermions is the operator 
\be
{\cal O}_{4} = :\!\psi \partial_{\tau} \psi \partial_{\tau}^2 \psi \partial_{\tau}^3 \psi\!:
\ee
which is quartic in the free fermion field and has dimension 8.

\begin{figure}[H]
\centering
\includegraphics[scale=0.5]{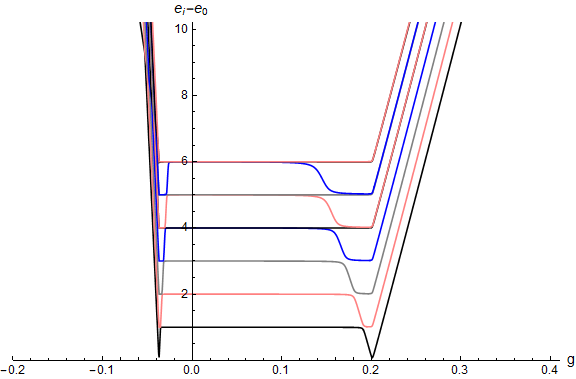}
\caption{\label{fig:your-figure}Perturbation of NS (13) fermions by the quartic term for both positive and negative coupling with $\Delta_{\rm max}=25.5$}
\label{fig17}
\end{figure}

Beginning at the IR point, we add a perturbation by the quartic term  to see if this can model  the flows beyond near the IR fixed point.  We consider perturbations of the form: 
\be \label{TplusO}
h^{\rm IR}= L_{0}^{\rm IR}  +g(T(0) \pm \eta{\cal O}_{4}(0))
\ee
where $\eta$ is some fixed constant.
  The operator ${\cal O}_{4}$ has very large matrix elements. To observe changes in the energy levels we found it convenient to set $\eta=5e^{-19}$. 
Looking  at the quartic interaction only, i.e. perturbing by $ \eta g{\cal O}_{4}(0)$ only, 
we get the following remarkably ordered picture (Fig. \ref{fig17}) in which the energy gaps $ e_{i}-e_{0}$ are presented against $g$.  
The behaviour is peculiar with almost all states dropping by one level at a very distinct value of  the coupling, which is then followed by an almost uniform linear regime.

Next, we look at the normalised energy gaps $\Delta e_{i}^{n}$ for different choices of sign   in (\ref{TplusO}) with $\Delta_{\rm max}=25.5$ corresponding to a state space of dimension 688.  
Here the spectator boundary condition is $(+)$ and the initial flowing boundary condition is   $(-)$.
\begin{figure}[H]
   \begin{subfigure}[b]{0.47\textwidth}
    \includegraphics[width=\textwidth]{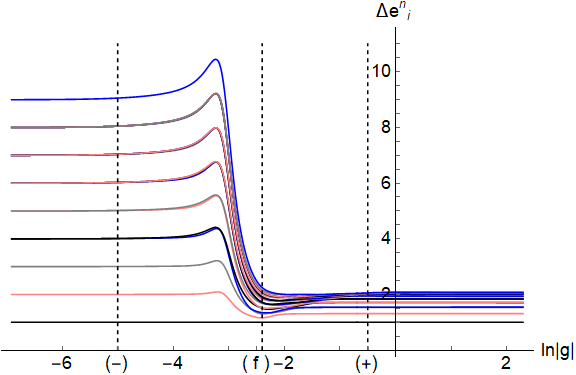}
    \caption{Perturbation of NS  fermions by $T$ and the quartic operator with positive $g$   and $+\eta$ chosen in (\ref{TplusO})}
    \label{fig:1}
  \end{subfigure}
  \qquad	
 \begin{subfigure}[b]{0.47\textwidth}
    \includegraphics[width=\textwidth]{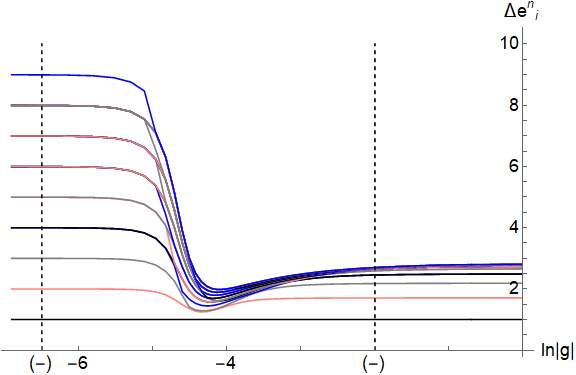}
    \caption{Perturbation of NS  fermions by $T$ and the quartic operator with positive $g$   and $-\eta$ chosen in (\ref{TplusO})}
    \label{fig:2}
  \end{subfigure}
  \caption{}
\end{figure}

The intermediate and the final multiplicities are different when we change the  sign in front of $\eta$.  For positive $\eta$ we indeed reach the UV fixed point describing Ramond fermions as one might expect.  For negative $\eta$ it is difficult to discern what happens to the multiplicities in this intermediate regime where it does not look as though the UV fixed point is reached.  The asymptotic multiplicities, however,  are easy to read off in both cases. With the plus sign in front of $\eta$ 
starting with the $(-)$ boundary condition we arrive at the $(+)$ one that corresponds to the end point of the $T$-perturbation flow (\ref{reverse2}). For the minus sign in front of $\eta$ the end multiplicities  are the same as  at the 
starting boundary condition. We get the same result for choosing the spectator to be $(-)$ that is the same as the initial 
 boundary condition. Hence we see that  this particular irrelevant perturbation, given by  (\ref{TplusO}) 
with the minus sign,  behaves like the end of the flow we observed in the 
flows beyond examples discussed above\footnote{The results are also similar whether we take positive or negative g.}.





\subsection{A maverick: Ising model with free spectator}\label{beyond5}
In section \ref{beyond_examples} we reviewed a number of examples of flows beyond the fixed point.  Locality would imply that the qualitative behaviour of these flows is independent of the choice of spectator boundary condition and indeed this is the case in the Tricritical Ising model.  In the Ising model the reverse flows are observed in the Ramond sector.  
For the free spectator we start with two primary towers present in the UV theory: the identity and the $\epsilon$-tower  
described by NS fermions. The multiplicities for low lying states up to level 8 are shown in Table \ref{expectedMultiplicites}.

\begin{figure}[H]\label{maverickPlots}
  \begin{subfigure}[b]{0.47\textwidth}
    \includegraphics[width=\textwidth]{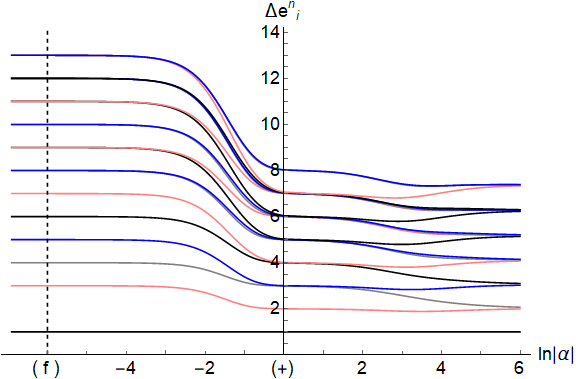}
    \caption{NS Sector.  TCSA normalised energy gaps with $\Delta_{\rm max}=31.5$ against ${\rm ln|\alpha|}$}
    \label{fig:1}
  \end{subfigure}
  \qquad	
  \begin{subfigure}[b]{0.47\textwidth}
    \includegraphics[width=\textwidth]{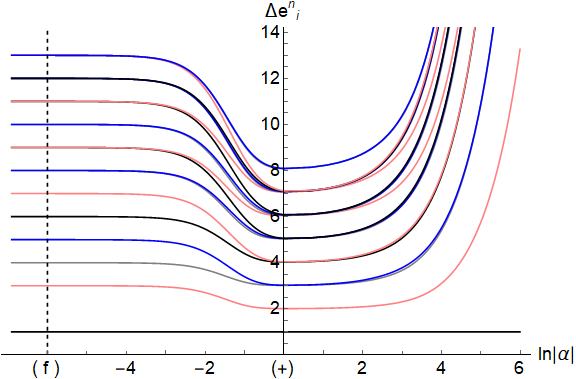}
    \caption{NS Sector. TCSA normalised energy gaps with $\Delta_{\rm max}=32$ against ${\rm ln|\alpha|}$}
    \label{fig:2}
  \end{subfigure}
  \caption{} \label{NS_maverik_plots}
\end{figure}

\begin{table}[H]
\centering
\resizebox{\textwidth}{!}{
\begin{tabular}{|l|l|l|l|l|l|l|l|l|l|l|l|l|l|l|l|l|l|}
\multicolumn{10}{l}{
}
 \\ \hline
 \textbf{$\Delta$} & 0 & 1/2 & 1 & 3/2 & 2 & 5/2 & 3 & 7/2 & 4 & 9/2 & 5 & 11/2 & 6 & 13/2 & 7 & 15/2 & 8 \\ \hline
$\mathbb{I}$ - tower    & 1 & 0   & 0 & 0   & 1 & 0   & 1 & 0   & 2 & 0   & 2 & 0    & 3 & 0    & 3 & 0    & 5 \\ \hline
$\epsilon$  - tower   & 0 & 1   & 0 & 1   & 0 & 1   & 0 & 1   & 0 & 2   & 0 & 2    & 0 & 3    & 0 & 4    & 0 \\ \hline
$\mathbb{I}$+$\epsilon$   towers     & 1 & 1   & 0 & 1   & 1 & 1   & 1 & 1   & 2 & 2   & 2 & 2    & 3 & 3    & 3 & 4    & 5 \\ \hline
\end{tabular}
}
\caption{Multiplicities of NS fermions} 
\label{expectedMultiplicites}

\end{table}
For a reversed flow we would expect the asymptotic multiplicities to  be the same as the starting ones -- given by both Virasoro towers. 
The numerical results plotted on Figure \ref{NS_maverik_plots} do not show this.  
Whilst we see that the IR fixed point is reached, with the correct multiplicities, there is a further rearrangement of the spectrum in the asymptotic regime where the multiplicities do not correspond to a Cardy boundary condition.


Moreover the asymptotic spectra depend on the truncation scheme, falling in two distinct patterns depending on whether 
$\Delta_{\rm max}$ is  integer or half-integer. In particular the asymptotic vacuum state becomes doubly degenerate for integer $\Delta_{\rm max}$ while it is unique for half-integer  $\Delta_{\rm max}$.  Whilst the flow certainly looks `deliberate', any mechanism for understanding this behaviour is as yet unexplained.  

The low lying asymptotic spectrum, although not described by a fixed point, can still be described in terms of the NS Virasoro 
towers\footnote{We are grateful to G. Watts for this observation.}. For half integer $\Delta_{\rm max}$ one should 
shift the weights in the $\epsilon$-tower by 1/2 and add this to the identity tower. For integer $\Delta_{\rm max}$ the asymptotic multiplicities can be obtained by shifting the weights in the 
$\epsilon$-tower by -1/2 and adding it to the identity tower. 
The resulting multiplicities are 
illustrated  in the two tables below.

\begin{table}[h]
\centering
\begin{tabular}{llllllllll}
                                                                      \\ \hline
\multicolumn{1}{|l|}{$\Delta$}               & \multicolumn{1}{l|}{0}   & \multicolumn{1}{l|}{1}   & \multicolumn{1}{l|}{2} & \multicolumn{1}{l|}{3} & \multicolumn{1}{l|}{4} & \multicolumn{1}{l|}{5} & \multicolumn{1}{l|}{6} & \multicolumn{1}{l|}{7} & \multicolumn{1}{l|}{8} \\ \hline
\multicolumn{1}{|l|}{$\mathbb{I}$ - tower} & \multicolumn{1}{l|}{1} & \multicolumn{1}{l|}{0}   & \multicolumn{1}{l|}{1} & \multicolumn{1}{l|}{1} & \multicolumn{1}{l|}{2} & \multicolumn{1}{l|}{2} & \multicolumn{1}{l|}{3} & \multicolumn{1}{l|}{3} & \multicolumn{1}{l|}{5} \\ \hline
\multicolumn{1}{|l|}{$\epsilon$ - tower with weights shifted by  +1/2}    & \multicolumn{1}{l|}{0}    & \multicolumn{1}{l|}{1} & \multicolumn{1}{l|}{1} & \multicolumn{1}{l|}{1} & \multicolumn{1}{l|}{1} & \multicolumn{1}{l|}{2} & \multicolumn{1}{l|}{2} & \multicolumn{1}{l|}{3} & \multicolumn{1}{l|}{4} \\ \hline
\multicolumn{1}{|l|}{TCSA asymptotic multiplicities}      & \multicolumn{1}{l|}{1} & \multicolumn{1}{l|}{1}   & \multicolumn{1}{l|}{2} & \multicolumn{1}{l|}{2} & \multicolumn{1}{l|}{3} & \multicolumn{1}{l|}{4} & \multicolumn{1}{l|}{5} & \multicolumn{1}{l|}{6} & \multicolumn{1}{l|}{9} \\ \hline
\end{tabular}
\caption{Observed multiplicities of the large coupling fixed point with half-integer $\Delta_{\rm max}$}
\label{halfIntTable}
\end{table}

\begin{table}[H]
\centering
\begin{tabular}{llllllllll}
                                                                                                       \\ \hline
\multicolumn{1}{|l|}{$\Delta$}              & \multicolumn{1}{l|}{0}   & \multicolumn{1}{l|}{1} & \multicolumn{1}{l|}{2} & \multicolumn{1}{l|}{3} & \multicolumn{1}{l|}{4} & \multicolumn{1}{l|}{5} & \multicolumn{1}{l|}{6} & \multicolumn{1}{l|}{7} & \multicolumn{1}{l|}{8}  \\ \hline
\multicolumn{1}{|l|}{$\mathbb{I}$ - tower} & \multicolumn{1}{l|}{1} & \multicolumn{1}{l|}{0} & \multicolumn{1}{l|}{1} & \multicolumn{1}{l|}{1} & \multicolumn{1}{l|}{2} & \multicolumn{1}{l|}{2} & \multicolumn{1}{l|}{3} & \multicolumn{1}{l|}{3} & \multicolumn{1}{l|}{5}  \\ \hline
\multicolumn{1}{|l|}{$\epsilon$ - tower with weights shifted by  -1/2}   & \multicolumn{1}{l|}{1} & \multicolumn{1}{l|}{1} & \multicolumn{1}{l|}{1} & \multicolumn{1}{l|}{1} & \multicolumn{1}{l|}{2} & \multicolumn{1}{l|}{2} & \multicolumn{1}{l|}{3} & \multicolumn{1}{l|}{4} & \multicolumn{1}{l|}{5}  \\ \hline
\multicolumn{1}{|l|}{TCSA asymptotic multiplicities}     & \multicolumn{1}{l|}{2} & \multicolumn{1}{l|}{1} & \multicolumn{1}{l|}{2} & \multicolumn{1}{l|}{2} & \multicolumn{1}{l|}{4} & \multicolumn{1}{l|}{4} & \multicolumn{1}{l|}{6} & \multicolumn{1}{l|}{7} & \multicolumn{1}{l|}{10} \\ \hline
\end{tabular}
\caption{Observed multiplicities of the large coupling fixed point with integer $\Delta_{\rm max}$}
\end{table}


This pattern of shifted towers leads us to think that the important characteristic of the truncation scheme is the 
difference between $\Delta_{\rm max}^{1}$ -- the maximal conformal weight in the truncated identity tower, 
 and $\Delta_{\rm max}^{\epsilon}$ -- the maximal conformal weight in the truncated $\epsilon$- tower. 
The difference $\Delta_{\rm max}^{\epsilon}-\Delta_{\rm max}^{1}$ is $-1/2$ when $\Delta_{\rm max}$ is integer and 
$+1/2$ when $\Delta_{\rm max}$ is half integer. We  test this pattern further using generalised truncation schemes  
(\ref{modif_scheme}) in which 
we use arbitrary positive integers $N^{1}_{\rm max}$, $N^{\epsilon}_{\rm max}$ to truncate the descendants in each 
tower. For such schemes we introduce a parameter
\be 
d=|\Delta_{\rm max}^{\epsilon}-\Delta_{\rm max}^{1}|=|1/2+N^{\epsilon}_{\rm max} -  N^{1}_{\rm max}| \, .
\ee
Using schemes with different values of $d$  we checked numerically that the asymptotic spectrum is 
described by shifting the weights of the tower with the larger value of $\Delta_{\rm max}^{i}$ by $d$ units up. 
For illustration in Table  \ref{d-shifts_table} we give the asymptotic weights corresponding to the scheme with $N^{1}_{\rm max}=30$ and 
$N^{\epsilon}_{\rm max}=31$ that has $d=3/2$. 
\begin{table}[h]
\centering
\begin{tabular}{llllllllll}
\hline
\multicolumn{1}{|l|}{$\Delta$}               & \multicolumn{1}{l|}{0}   & \multicolumn{1}{l|}{1} & \multicolumn{1}{l|}{2}   & \multicolumn{1}{l|}{3} & \multicolumn{1}{l|}{4} & \multicolumn{1}{l|}{5} & \multicolumn{1}{l|}{6} & \multicolumn{1}{l|}{7} & \multicolumn{1}{l|}{8} \\ \hline
\multicolumn{1}{|l|}{$\mathbb{I}$ - tower} & \multicolumn{1}{l|}{1} & \multicolumn{1}{l|}{0} & \multicolumn{1}{l|}{1}   & \multicolumn{1}{l|}{1} & \multicolumn{1}{l|}{2} & \multicolumn{1}{l|}{2} & \multicolumn{1}{l|}{3} & \multicolumn{1}{l|}{3} & \multicolumn{1}{l|}{5} \\ \hline
\multicolumn{1}{|l|}{$\epsilon$ -  tower with weights shifted by +3/2}    & \multicolumn{1}{l|}{0}    & \multicolumn{1}{l|}{0}  & \multicolumn{1}{l|}{1} & \multicolumn{1}{l|}{1} & \multicolumn{1}{l|}{1} & \multicolumn{1}{l|}{1} & \multicolumn{1}{l|}{2} & \multicolumn{1}{l|}{2} & \multicolumn{1}{l|}{3} \\ \hline
\multicolumn{1}{|l|}{TCSA asymptotic multiplicities}      & \multicolumn{1}{l|}{1} & \multicolumn{1}{l|}{0} & \multicolumn{1}{l|}{2}   & \multicolumn{1}{l|}{2} & \multicolumn{1}{l|}{3} & \multicolumn{1}{l|}{3} & \multicolumn{1}{l|}{5} & \multicolumn{1}{l|}{5} & \multicolumn{1}{l|}{8} \\ \hline
\end{tabular}
\caption{observed multiplicities of the large coupling fixed point with $\Delta_{\rm max}=31.5,\text{ } \Delta^{1} = 30,\text{ } \Delta^{\epsilon}= 31.5 \text{ and } d=3/2$} \label{d-shifts_table}
\end{table}

\begin{figure}[H]
\centering
\includegraphics[scale=0.5]{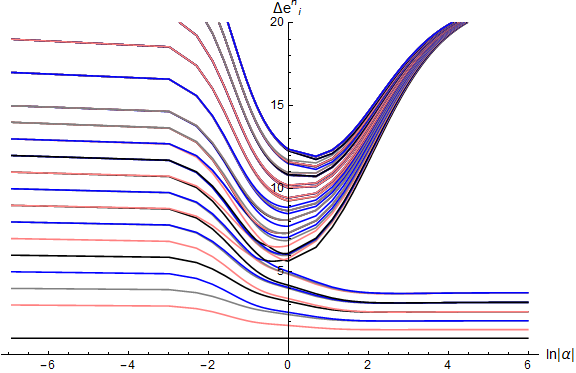}
\caption{$\Delta e^{n}_{i}$ against ${\rm ln}|\alpha|$ for free spectator, positive coupling and  $N^{1}_{\rm \max}=7$, $N^{\epsilon}_{\rm max}=21$}
\label{fig:your-figure20}
\end{figure}

For large values of the splitting parameter $d$ the two towers become completely decoupled (as far as the low energy states are concerned).  
The plot on Fig. \ref{fig:your-figure20} shows the normalised energy gaps for the scheme with 
$N^{1}_{\rm max}=7$ and $N^{\epsilon}_{\rm max}=21$ with $d=14.5$.

 In the asymptotic regime the flow is dominated by the interaction matrix.  
The perturbing term in (\ref{NS Ham}) has a block structure with interacting terms coming only from the mixing of states between the two towers.  
Increasing the split between the two towers we remove  matrix elements from the interaction. The multiplicities of the low lying spectrum match those of the smallest tower.  The second tower is essentially decoupled from this low lying spectrum.  
The tight band of levels that are lifted in the asymptotic regime  visible on the plot above corresponds 
to the zero eigenvalue of the interaction matrix. Its multiplicity is given by the difference between dimensions of the two truncated towers.

 The split in energies between the two towers can be expected based on   renormalisation group analysis of 
 truncation effects introduced in \cite{Gerard2}. As shown in \cite{Gerard2} whenever a self coupling is present 
 perturbatively, via a non-vanishing three-point function of the perturbing operator, as is the case in TIM, 
 it gives the dominant 
 truncation effect in the form of a coupling constant running with $\Delta_{\rm max}$. In the Ising model 
 such a self coupling is absent and   it was shown  in \cite{Gerard3} that the leading truncation effect 
 can be described via  a coupling to a non-local operator:  $(-1)^{N}$,  where $N$ is the free fermion number 
 \be
 N=\sum_{n=0}^{\infty}a^{\dagger}_{n+1/2}a_{n+1/2}\, .
 \ee
It appears to be important to understand in general when  the non-local operators are important for  TCSA 
truncation effects. We hope to return to this question in future work. 

\subsection{Different truncation methods for TIM}\label{beyond4}
\begin{figure}[h!]
\centering
\includegraphics[scale=0.5]{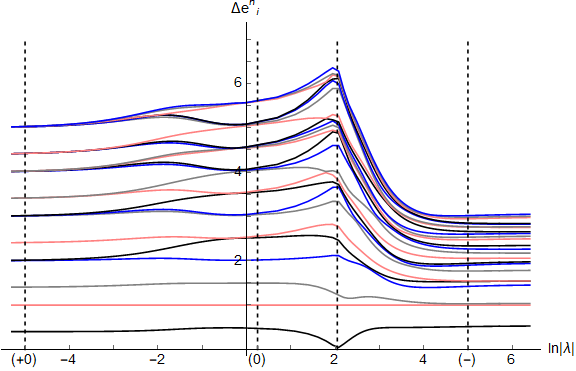}
\caption{\label{fig:your-figure}Normalised Energies (wrt E$_2$) in the perturbed TIM with $(d)$ spectator for negative coupling.  Respective truncation levels $N^{\sigma}=20$, $N^{\sigma^{\prime}}=18$}
\label{TIM_diff_schemes}
\end{figure}

Whenever we have more than one primary tower in the UV spectrum we can apply the truncation schemes 
   (\ref{modif_scheme}). In the tricritical Ising model for the UV boundary condition $(0+)$ and the spectator $(d)$, the state space contains two Virasoro towers corresponding to the primaries $\sigma$ and $\sigma^{\prime}$.
For the truncation schemes with $N^{\sigma}=20 $ and $N^{\sigma'}=18,19$ we find that for positive coupling the flow gets close to the $(+)$ fixed point and that the flows beyond are susceptible to these changes in truncation scheme, with, in particular, some examples showing rearrangements in the multiplicities which do not correspond to a known fixed point.
 For negative coupling we obtain a picture qualitatively different from the simple truncation scheme. 
 Past the first IR fixed point $(0)$ we obtain a flow beyond to multiplicities that may not correspond to a known fixed point (see the plot below where the vacuum becomes degenerate).  However, past this region the multiplicities rearrange further with the 3 low lying levels being the same as that describing the $(-)$ boundary condition that is the left end of the cascade of TCSA flows. 
 On Figure \ref{TIM_diff_schemes} the reader can see a plot of energy gaps normalised by dividing them by $e_{2}-e_{0}$. 


  The effects of  increasing the split in truncation levels between the two towers   are similar to the Ising model in the NS sector.  However, some important differences do remain.
  In particular, while for some truncation schemes we pass through a point that cannot be identified as a fixed point, asymptotically we always recover the multiplicities of the $(-)$ fixed point for negative coupling.  For positive coupling, the multiplicities of the $(+)$ fixed point are not always reached. For the simple truncation scheme governed by $\Delta_{\rm max}$ we always flow through identifiable fixed points. The difference with the Ising model can be explained 
  by the absence in the latter of the self-coupling of the $\epsilon$ operator that would be responsible for the leading truncation effects \cite{Gerard2}.



\section{Concluding remarks}\label{conclusions_sec}
In this section we attempt to summarise our investigations and point to various loose ends. 
In section  \ref{beyond_general} we offered a tentative explanation for why the leading truncation error is linear. It is desirable to understand this better, confirming or disproving that scenario.
It would also be interesting to understand the subleading corrections and improve upon the numerical accuracy of the method. 

In section \ref{Approach_section} we studied numerically how TCSA behaves near the IR fixed point. We identified the leading 
truncation error in that region as a certain linear function. We designed two methods 
that subtract the leading error and extract a numerical approximation to the leading infrared exponent 
that is the dimension of the leading irrelevant operator along which the RG trajectory approaches the fixed point.

In section \ref{FlowsBeyond_section} we studied the behaviour of TCSA energy levels past the closest approach to the IR fixed point. 
There we observed, confirming prior work by other people,  flows beyond where the energy levels 
rearrange their multiplicities passing through several discernible regimes. We investigated how these flows depend on the 
choice of the spectator boundary condition and on the choice of the truncation scheme. We presented examples of 
3  types of behaviour: when the flow beyond arrives to a known fixed point, reversing a known RG flow; 
when the flow beyond arrives to a set of multiplicities not corresponding to any fixed point; when there is no flow beyond at all. We found that one scenario can be replaced by another when we change the truncation scheme and/or the spectator boundary condition. Our working hypothesis, discussed in section  \ref{beyond_general}, is that the TCSA behaviour near the IR fixed point can be described by the 
continuum theory with a number of irrelevant {\it local} operators switched on. Based on this  we described a possible scenario of how the flow beyond may originate, starting at the IR fixed point, via the leading irrelevant perturbation with the reversed sign of the coupling.  The behaviour of the TCSA spectrum past the IR fixed point then depends heavily on the truncation scheme used. We were able to model the qualitative behaviour of TCSA flows beyond 
by adding higher dimension irrelevant perturbations to the leading one. The assumption of locality of an effective description remains essential in describing flows that pass or end at a physical fixed point. In some cases, like the boundary magnetic field perturbation in the Ising model with a free spectator, we observed a flow to an unphysical set of multiplicities. The fact that this depends on a choice of spectator boundary condition suggests that some non-local operators may be needed to describe TCSA corrections in such cases. This is corroborated by a perturbative calculation in that model \cite{Gerard3}. It is important for future progress to understand theoretically in which cases the non-local operators are important and how to take them into account, or how to modify the truncation scheme to suppress their effects.

\begin{center}
{\bf \Large Acknowledgments }
\end{center}
The authors are grateful to Gerard Watts for many stimulating discussions. A.K. wants to thank G\'abor Tak\'acs for 
sharing his knowledge of TCSA. 


\appendix

\renewcommand{\theequation}{\Alph{section}.\arabic{equation}}
\setcounter{equation}{0}
\section{Dimensions of the truncated space}
\subsection{The Ising model}\label{App1}
Tables \ref{tbl:tableA1} and \ref{tableA2} contain  the dimensions of truncated physical space depending on $\Delta_{\rm max} $ for the fixed spectator and free spectators respectively.

\begin{table}[h!]
\centering
\begin{tabular}{|p{4cm}|c|c|c|c|c|c|c|c|c|c|}
\hline
$\Delta_{\rm max}$ &17 &19 &21 &24 & 26 &29 & 32 & 34 & 36 & 40\\
\hline
 dimension of truncated  space& 207 &307 &447 & 762 &1069 &1739 &2765 &3725 &4978 & 8697 \\
 \hline
 \end{tabular}
\caption{Dimensions of truncated spaces  of Ramond fermions} \label{tbl:tableA1}
\end{table}

\begin{table}[h!]
\centering
\begin{tabular}{|p{4cm}|c|c|c|c|c|c|c|c|c|c|}
\hline
$\Delta_{\rm max} $ &19.5 &20 &24.5 &25 &29.5 &30 & 34.5 & 35 & 37.5 &38\\
\hline
dimension of truncated space with even number of oscillators &214 &260 &534 &632 &1217 &1426 &2611 & 3019 & 4020 & 4629\\
\hline
dimension of truncated space with odd number of oscillators &236 &236 &581 &581 &1317 &1317 &2809 & 2809 & 4315 & 4315 \\
\hline 
 total dimension of truncated  space&450 &496 &1115 &1213 &2534 &2743 &5420 &5828 & 8335 &8944 \\
 \hline
  \end{tabular}
\caption{Dimensions of truncated spaces  of NS fermions} \label{tableA2}
\end{table}

\subsection{The tricritical Ising model}\label{App2}
Table \ref{tbl:tableA3} contains  the dimensions of truncated physical space depending on $\Delta_{\rm max} $ for the $(d)$ spectator in TIM. 

\begin{table}[h!]
\centering
\begin{tabular}{|p{4cm}|c|c|c|c|c|}
\hline
$\Delta_{\rm max}$ &16&17 &18&19&20\\
\hline
 dimension of truncated  space& 848&1082 &1373 &1731& 2170\\
 \hline
 \end{tabular}
\caption{Dimensions of truncated spaces in the tricritical Ising model with (d) spectator} \label{tbl:tableA3}
\end{table}

\end{document}